\documentclass[aps,pre,floatfix,reprint,superscriptaddress,nobibnotes,showkeys]{revtex4-2}
\usepackage[latin9]{inputenc}
\usepackage{float}
\usepackage[margin=3cm]{geometry}
\usepackage{hyperref}
\hypersetup{pdfauthor={Ata Mesgarnejad, Northeastern University},
	    	colorlinks,
	    	pdfcreator={pdflatex},
	    	unicode=false,
	    	pdftoolbar=false,
	    	pdfmenubar=true,
	   		pdffitwindow=true,
	   		pdfnewwindow=true,
	    	linkcolor=red,
	    	citecolor=red,
	    	filecolor=black,
	    	urlcolor=red,
	    	}
\usepackage{mathrsfs}
\usepackage{amsthm}
\usepackage{amsmath}
\usepackage{amssymb}
\usepackage{graphicx}
\usepackage{pbox}
\usepackage{esint}
\usepackage{nomencl}
\usepackage{diagbox,ragged2e}
\usepackage{algpseudocode}
\makeatletter
\usepackage[normalem]{ulem}

\floatstyle{ruled}

\newfloat{algorithm}{tbp}{loa}
\floatname{algorithm}{Algorithm}
\theoremstyle{plain}

\theoremstyle{definition}

\theoremstyle{remark}

\usepackage{amsxtra}
\usepackage{mathrsfs}
\usepackage{amsthm}
\usepackage{pdfsync}
\usepackage{booktabs}
\usepackage[version=4]{mhchem}
\usepackage{xargs}                      

\newcommand{\ie}{\emph{i.e., }}
\newcommand{\eg}{\emph{e.g., }}

\newcommand{\C}{\mathcal{C}}
\newcommand{\A}{\mathcal{A}}
\newcommand{\F}{\mathcal{F}}
\newcommand{\ub}{u}
\newcommand{\xb}{x}
\newcommand{\micron}{\mu\mathrm{m}}
\newcommand{\argmax}{\mathrm{argmax}}
\newcommand{\mm}{\mathrm{mm}}
\newcommand{\vol}{\mathrm{Vol}\%}
\newcommand{\Fig}{Fig.}
\usepackage{float}
\renewcommand{\figurename}{FIG}
\usepackage{xcolor}
\newcommandx{\AM}[1]{\textcolor{blue}{#1}}
\usepackage[flushleft]{threeparttable}     

\begin{document}
\title{Crack Path Selection in Orientationally Ordered Composites}
\author{A.~Mesgarnejad}
\affiliation{Center for Inter-disciplinary Research on Complex Systems, Department of Physics, Northeastern University, Boston, MA. 02115, U.S.A.}
\author{C.~Pan}
\author{R.M.~Erb}
\affiliation{Department of Mechanical And Industrial Engineering, Northeastern University, Boston, MA. 02115, U.S.A.}
\author{S.J.~Shefelbine}
\affiliation{Department of Mechanical And Industrial Engineering, Northeastern University, Boston, MA. 02115, U.S.A.}
\affiliation{Department of Bioengineering, Northeastern University, Boston, MA. 02115, U.S.A.}
\author{A.~Karma}
\email{a.karma@northeastern.edu}
\affiliation{Center for Inter-disciplinary Research on Complex Systems, Department of Physics, Northeastern University, Boston, MA. 02115, U.S.A.}
\vspace{-10em}
\begin{abstract}
While cracks in isotropic homogeneous materials propagate straight, perpendicularly to the tensile axis cracks in natural and synthetic composites deflect from a straight path, often increasing the toughness of the material. 
Here we combine experiments and simulations to identify materials properties that predict whether cracks propagate straight or kink on a macroscale larger than the composite microstructure. 
Those properties include the anisotropy of the fracture energy, which we vary several folds by increasing the volume fraction of orientationally ordered alumina platelets inside a polymer matrix, and a microstructure-dependent process zone size that is found to modulate the additional stabilizing or destabilizing effect of the non-singular stress acting parallel to the crack. 
Those properties predict the existence of an anisotropy threshold for crack kinking and explain the surprisingly strong dependence of this threshold on sample geometry and load distribution.
\end{abstract}

\maketitle

\section{Introduction}\label{sec:introduction}

In natural and synthetic composites consisting of hard particles within a soft matrix, crack path prediction is a complex, intrinsically multiscale, problem. Accurate prediction of crack paths, especially the existence of kinking, provides insight into the properties of fracture toughness and strength. Therefore, understanding how cracks propagate at the scale of the hard particles (microscale) and at the scale much larger than the particles (macroscale) are both essential. On the microscale, cracks can either penetrate or be deflected by the hard particles depending on the elastic and fracture properties of the two phases~\cite{he1989kinking,ahn1998criteria,Tankasala:2017}. 
Such microscale deflection has been hypothesized to provide an apparent toughening mechanism in both natural~\cite{barthelat2007experimental,ritchie2011conflicts,sen2011structural,dimas2013tough} and biomimetic~\cite{dimas2013tough,mirkhalaf2014overcoming} composites by greatly increasing the fracture surface area and the required energy to fracture the material. 
Crack deflection at the microscale is considered well understood.
Microscale deflection can potentially, but not always, lead to macroscale crack deflection even in pure tensile (mode-I) loading configurations which has been studied extensively in both natural~\cite{Koester:2008,carriero2014tough} and biomimetic composites~\cite{dimas2013tough,suksangpanya2018crack}. 
For example macroscale kinking occurs in healthy bone for cracks perpendicular to the collagen fiber direction while straight crack propagation has been seen in pathological bone exhibiting disordered collagen fibers~\cite{carriero2014tough}. 
Other micro-structurally ordered natural composites that have impressive fracture toughness and macroscale crack kinking include seashells~\cite{Meyers:2008,Huang:2013,Sakhavand:2015}, wood~\cite{Schachner:2000,Amada:2001}, dental enamel~\cite{Bechtle:2010}, and rock~\cite{Hoagland:1973,Chandler:2016}. 
While these observations suggest that macroscale crack kinking may result from microscale alignment, other structural heterogeneities such as modulus variations and porosities in natural composites may also contribute to kinking~\cite{Koester:2008}.

In synthetic composites, in which other heterogeneities can be minimized, micro-sized particles or fibers are added to concrete~\cite{Gopalaratnam:1995}, ceramics~\cite{Yang:2008,Liu:2013a}, and polymers~\cite{Kim:2008a} to increase toughness. 
The particles or  fibers in synthetic composites are often not arranged at the micro-structural level as they are in natural materials and can show crack kinking~\cite{Koester:2008,Huang:2013} as well as straight crack propagation~\cite{carriero2014tough}.  
When the micro-structure is well-aligned, such as in freeze-casted nacre-like alumina samples, and subjected to notched three-point bending, crack kinking occurs when the crack direction is perpendicular to the microstructure orientation~\cite{Bouville:2014}. 
3D printed composites have also shown crack kinking both with aligned microstructure~\cite{Martin:2015} and macrostructures~\cite{dimas2013tough}.

Previous experiments on natural and synthetic composites have explored the influence of anisotropy on crack path, where they have found that crack deflection depends on fracture toughness anisotropy, the direction of the crack relative to the aligned microstructure, and the volume fraction of the particles~\cite{Keck:2016}. 
These studies have developed regression models to predict the kink angle~\cite{Keck:2016}, estimated a critical anisotropy ratio for crack kinking~\cite{Judt:2019}, and developed a geometric adjustment factor for samples with fiber direction perpendicular to the crack direction~\cite{Keck:2016}.  
In the present work, we use a rather unique composite system that allows us to vary the fracture energy anisotropy several fold while keeping the elastic properties nearly isotropic. Therefore, we are able to investigate the effect of the fracture energy anisotropy on crack paths on a macroscale (i.e. with both the fracture energy and the crack paths measured on a sample scale much larger than the composite microstructure). This is achieved by leveraging magnetic alignment to produce highly oriented and homogeneous alumina microstructures within photocurable polymers. By conducting Mode I tensile testing of notched specimens with these composites in different geometries that promote or suppress kinking, we are able to quantify experimentally the fracture energy anisotropy and, at the same time, to demonstrate a surprisingly strong effect of sample geometry on crack kinking behavior. 

Furthermore, to explain our experimental findings, we 
use the phase-field approach for fracture~\cite{Francfort:1998,Karma:2001a,Bourdin:2008a}. This method
has been validated by theoretical analyses~\cite{Francfort:1998,Hakim:2009} and comparisons with observed crack paths in benchmark geometries~\cite{Mesgarnejad:2015}. It has been used to model a wide range of fracture phenomena in diverse applications including thin-film fracture~\cite{Mesgarnejad:2013}, thermal fracture \cite{Bourdin:2014a}, mixed mode fracture~\cite{Chen:2015b}, chemo-mechanical fracture~\cite{Miehe:2015,Zuo:2015,Klinsmann:2016,Klinsmann:2016a}, dynamic fracture~\cite{Chen:2015b,Chen:2017a,Lubomirsky:2018}, fracture in colloidal systems~\cite{Peco:2019}, as well as ductile fracture~\cite{Mozaffari:2015,Ambati:2015,Borden:2016,Alessi:2017} and fatigue crack growth~\cite{Alessi:2018a,Carrara:2018,Mesgarnejad:2019}. Directly relevant to the present study, the phase-field method has also been used to model brittle fracture with an anisotropic fracture energy \cite{hakim2005crack,li2015phase,Hakim:2009} and fracture of composites at micro~\cite{murali2011role,Msekh:2016,Wei:2019} and macro~\cite{Doan:2016a,Hirshikesh:2019} scales. Here, we perform phase-field simulations that demonstrate the existence of a transition from straight to kinked crack propagation on a macro scale with increasing magnitude of the fracture energy anisotropy in good quantitative agreement with experimental findings. Simulations also reproduce the
surprisingly strong effect of sample geometry on crack kinking behavior for values of the process zone size in the phase-field model estimated from experimentally measured mechanical properties. We explain quantitatively this effect 
in terms of the non-singular T-stress acting parallel to the crack. 
While the T-stress has been found to influence crack path selection in isotropic media such as PMMA \cite{Ayatollahi:2016}, its effect has so far been neglected in theoretical studies of crack paths in anisotropic media
\cite{hakim2005crack,li2015phase,Hakim:2009}. Here we show that the dependence of crack kinking behavior on both the anisotropy of the fracture energy and sample geometry can be quantitatively predicted using existing analytical predictions for the energy release rate $G$ at the tip of a short kinked extension of a pre-existent crack in mode I loading, which take into account the contributions of both the singular stresses and the T-stress \cite{Amestoy:1992}. 

When the T-stress is neglected, as in previous studies \cite{hakim2005crack,Hakim:2009}, $G$ depends only on the mode I stress intensity factor $K_I$ and the kink angle $\theta$, which together determine the mode I and mode II stress intensity factors at the tip of the kinked crack, $k_{I}$ and $k_{II}$ respectively, and hence the energy release rate $G=(k_I^2+k_{II}^2 )/E$, where $E$ is the elastic modulus. When the T-stress is included, $G$ depends additionally on the length $s$ of the kinked crack extension through the ratio $T\sqrt{s}/K_I$ of non-singular and singular stress-fields. Hence, $G$ depends additionally on both the sample geometry and loading configuration, which determine the magnitude of $T$, and the size $\xi$ of the process zone around the crack tip where linear elasticity breaks down, which sets a natural length scale for $s$. Remarkably, we find that computing $G$ with the simple choice $s=\xi$ predicts quantitatively well crack kinking behavior in the experiments, where $\xi$ is estimated from measured mechanical properties, and in the phase-field simulations where $\xi$ is a key input parameter together with the fracture energy anisotropy.
The combined experimental and numerical results provide a comprehensive understanding of the combined effects of fracture energy anisotropy and sample geometry on macroscale crack paths in anisotropic composites. 

\section{Materials and Methods}\label{sec:methods}

\subsection{Preparation of oriented Alumina-reinforced polymer matrix composites}

For the experiments, we used composites with varying volume fraction $f_v$ of uniformly dispersed micron-size alumina platelets embedded in a polymeric matrix that exhibited little plastic deformation. 
Platelets were dispersed within uncured polymeric resin and then forced to orient in a common plane by applying a magnetic field, thereby producing a composite with a fine microstructure with long-range orientational order. 
To produce the polymer matrix, we mixed two photocurable resins, \emph{EBECRYL} 230 urethane acrylate and isobornyl acrylate, with a weight ratio of 1:9 so as to raise the viscosity and prevent the sedimentation of particles, together with two photoinitiators (\emph{1-Hydroxycyclohexyl phenyl ketone} and \emph{Phenylbis phosphine oxide}, $1\,\mathrm{wt}\%$ each relative to the resin). 
To this, magnetized alumina particles were added at $0\textup{--}7\,\vol$ ($f_v=0-0.07$). 

To produce the magnetized Alumina particles, first $10\,\mathrm{g}$ of $7.5\,\micron$ alumina (Antaria, Australia) micro-platelets were dispersed in $400\,\mathrm{mL}$ of deionized water in an Erlenmeyer flask with a magnetic stirring bar stirring at $500\,\mathrm{rpm}$. 
The pH of the water was maintained as $7$ under room temperature to keep a positive charge on the alumina surface (isoelectric point at $\mathrm{pH}\approx9$). 
Separately, $375\,\mu\mathrm{L}$ of super paramagnetic iron oxide nanoparticles (SPIONS, EMG 705, Ferrotec, Nashua, NH) were diluted with $40\,\mathrm{mL}$ of deionized water. 
The diluted dispersion was added dropwise into the suspension with alumina particles. 
The negatively charged SPIONs electrostatically coated the positively charged alumina micro-platelets. 
Typically, the adsorption was complete within 24 hours as was determined when the supernatant liquid was transparent. The magnetized alumina was isolated through vacuum filtration in a Buchner filter. 
The particles were dried in an oven at $60\,^\circ\mathrm{C}$ for at least 12 hours and were stored in a desiccator chamber with a humidity below $10\%$.

\begin{figure}[ht!]
	\centering
	\includegraphics[width=1\columnwidth]{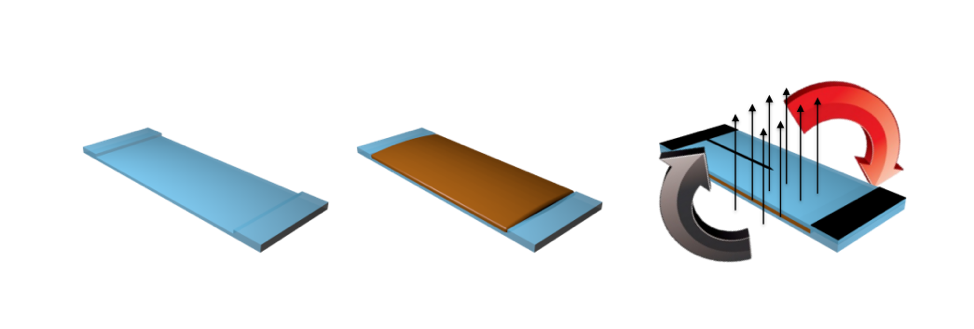}
	\caption[Composites' molding and curing process]{\small Composites fabrication: glass slide with two glass spacers at each ends (left), resin spread on the glass slide (middle), resin covered by the glass slide with photomask and placed in the oscillating field (right).}
	\label{fig:fabrication}
\end{figure}

Urethane (EBECRYL 230, Allnex) and isobornyl acrylate (IBA, Sigma Aldrich) were mixed with a weight ratio of 1:9. The two resins were made photo-curable by adding both 1-hydroxycyclohexyl phenyl and (Sigma Aldrich) phenylbis (2,4,6-trimethylbenzoyl) phosphine oxide (Sigma Aldrich) as photo initiators at a weight ratio of 1\% each. 
The resin was mixed with a spatula and sonicated for 1 min. Magnetized alumina at defined volume fractions was then added and dispersed with an ultrasonic probe (Sonifier 250, Branson) for 2 min with output power 2 and duty cycle of 20. The solution was transferred to sonicate bath for another 20 min sonication to further ensure that the magnetized alumina is homogeneously dispersed in the polymer resin.
The sonicated solution was then spread onto a glass slide by a transfer pipet and covered by another glass slide with a photomask to form a sandwich mold. The glass slide and the photomask were separated by $0.3\,\mathrm{mm}$ thick glass spacers (see \Fig~\ref{fig:fabrication}).

The mold was then placed into an oscillating magnetic field created by solenoids powered with current controllers that were themselves controlled by a {LabVIEW}~\cite{labView} program.  
The oscillating field consisted of a constant vertical field (Z-field) of 140 G and a dynamic horizontal field (X-field) that was a $1.5\,\mathrm{Hz}$ sinusoidal field with a peak field of 240 G. Due to the ultrahigh magnetic response, the magnetized alumina could be aligned within the X-Z plane in 10 seconds. 
After magnetized alumina was assembled, ultraviolet (UV) light emitted from a UV lamp (6 W, 365 nm) was applied 10 cm above the sample curing the unmasked area. Samples with different volume fraction of alumina micro-platelets required different curing time to obtain a testable thickness. Typically, exposure of $17\,\mathrm{s}$, $23\,\mathrm{s}$, $35\,\mathrm{s}$, $50\,\mathrm{s}$, $60\,\mathrm{s}$, $70\,\mathrm{s}$, $75\,\mathrm{s}$, and $90\,\mathrm{s}$ was required to cure $0\%$, $1\%$, $2\%$, $3\%$, $4\%$, $5\%$, $6\%$, $7\%$, and $10\%$ filled samples, respectively.  
Next, the mold was carefully peeled apart leaving the sample stuck to the glass slide with the photomask. The surface of the sample was cleaned by isopropanol (IPA, Sigma-Aldrich) and was then flipped over and subjected to the same UV light for 1 min of short post-curing before it was peeled off by a razor blade and transferred to a container. 
For further post-curing, the light of a digital light processing ({DLP}) projector placed 25 cm away from the sample was used to post cure each side of the sample for $15\,\mathrm{min}$. Afterwards, the notch position was marked $1\,\mm$ ahead of the tip before the sample was placed in an oven at $90\,^\circ\mathrm{C}$. 
The notch was made by a sharp razor blade after the sample was heated for $20\,\mathrm{min}$ and became soft. The notched sample was further heat treated in the oven at $90\,^\circ\mathrm{C}$ for another 2 hours to relax the residual stress that may have been created at the crack tip during notching . During the heat treatment, glass slides was used to cover the sample to prevent warping.

\subsection{Characterization of particle dispersion in experimental samples}\label{sec:spacing}

\begin{figure*}[htb!]
	\centering
	\includegraphics[width=2\columnwidth]{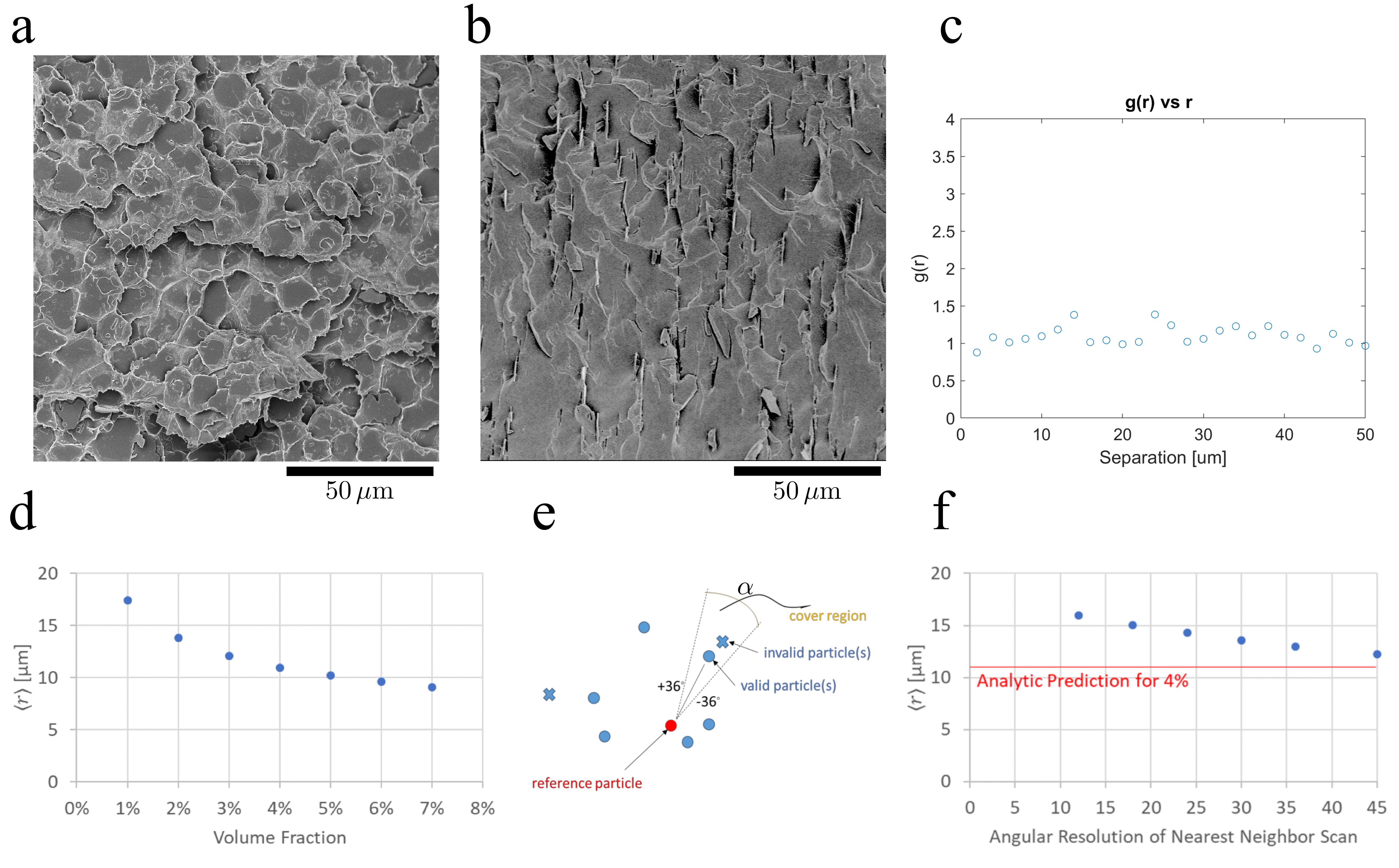}
	\caption[Characterization of particle spacing]{
	\textbf{a}-\textbf{b}, SEM cross-sections taken from fractured samples reinforced with 4\% alumina oriented $\parallel$ (\textbf{a}) and $\perp$ (\textbf{b}, obtained from freeze fracture to avoid kinking).
	\textbf{c}, Plot of the pair correlation (radial distribution) function, $g(r)$ vs. platelet separation  $r$ calculate using graphically identified platelets in a $\perp$ sample.
	\textbf{d}, Analytical prediction of nearest neighbor spacing using the relationship $\langle r\rangle\sim n^{-1/3}$. 
	\textbf{e}, Schematic of the algorithm for locating nearest neighbors with a certain angular resolution, $\alpha$, disregarding more distant particles. 
	\textbf{f}, Results of the custom algorithm showing a slight dependence on the angular resolution, but being in reasonable agreement with the analytical expression for a volume fraction of 4\%.}
	\label{fig:SEM-spacing}
\end{figure*}

To visualize the microstructure, fabricated composites were freeze fractured in liquid nitrogen to expose a cross-section without any plastic deformation. 
These cross-sections were observed in a scanning electron microscope (SEM) as shown in \Fig~\ref{fig:SEM-spacing}a and  \Fig~\ref{fig:SEM-spacing}b for samples with $\parallel$ ($0^\circ$)  and $\perp$ ($90^\circ$) platelet orientations, respectively.
Investigation of the microstructure shows that the ceramic platelets are homogeneously distributed with average nearest neighbor separation, $\langle r\rangle$, around $11\,\micron$. 
This inter-particle-spacing was established through analytic predictions and graphical analysis. 
We calculated the pair correlation (radial distribution) function, $g(r)$, for the platelets by graphically identifying platelet centers in the $90^\circ$ case using ImageJ~\cite{Rueden:2017} analysis software supplied by the NIH, shown in~\Fig~\ref{fig:SEM-spacing}c. 
The pair correlation indicates that there is almost no correlation between particle positions in long or short range, indicating that there is no clustering or ordering. 
In other words, the alumina is randomly and homogeneously distributed. It should be noted that $g(r)$ can therefore not be relied upon to establish nearest neighbor distances. 
We investigated nearest neighbor distance using the analytical expression for homogeneous particle suspensions 
$\langle r\rangle\simeq(V_p/f_v)^{1/3}=n^{-1/3}$ where $V_p$ is a single platelet's volume, and $n=f_v/V_p$ is the platelet number density.
To calculate the platelet number density, $n$, we assumed that the platelets were discs with volume $V_p=\pi a^2 t$, where the platelet diameter was $a=7.5\,\micron$ and thickness was $t=0.3\,\micron$. 
The analytically calculated nearest neighbor separation $\langle r\rangle$ is dependent on the platelet volume fraction as shown in ~\Fig~\ref{fig:SEM-spacing}d. 
For the 4\% volume fraction sample shown in~\Fig~\ref{fig:SEM-spacing}, the analytic equation predicts roughly $\langle r\rangle\sim11\,\micron$. To verify this prediction, and since we are unable to use $g(r)$, we employed a custom algorithm to calculate the average distance to the nearest neighbor from each particle over a variable angular resolution. This algorithm is depicted in ~\Fig~\ref{fig:SEM-spacing}e in which only the closest particle over a scan across an angle of $\alpha$ is considered. All of the closest particles are then averaged over $360^\circ$. 
This approach produces the results shown in~\Fig~\ref{fig:SEM-spacing}f for different angular resolutions of $\alpha$. To avoid edge effects, boundaries are made periodic by arraying particle positions in x and y. This algorithm indicates the separation between nearest neighbors is $\langle r\rangle\sim14\,\micron$ indicating a reasonable agreement with the analytical model.

\subsection{Mechanical testing procedures and crack path mapping}

\begin{figure}[ht!]
	\centering
	\includegraphics[width=\columnwidth]{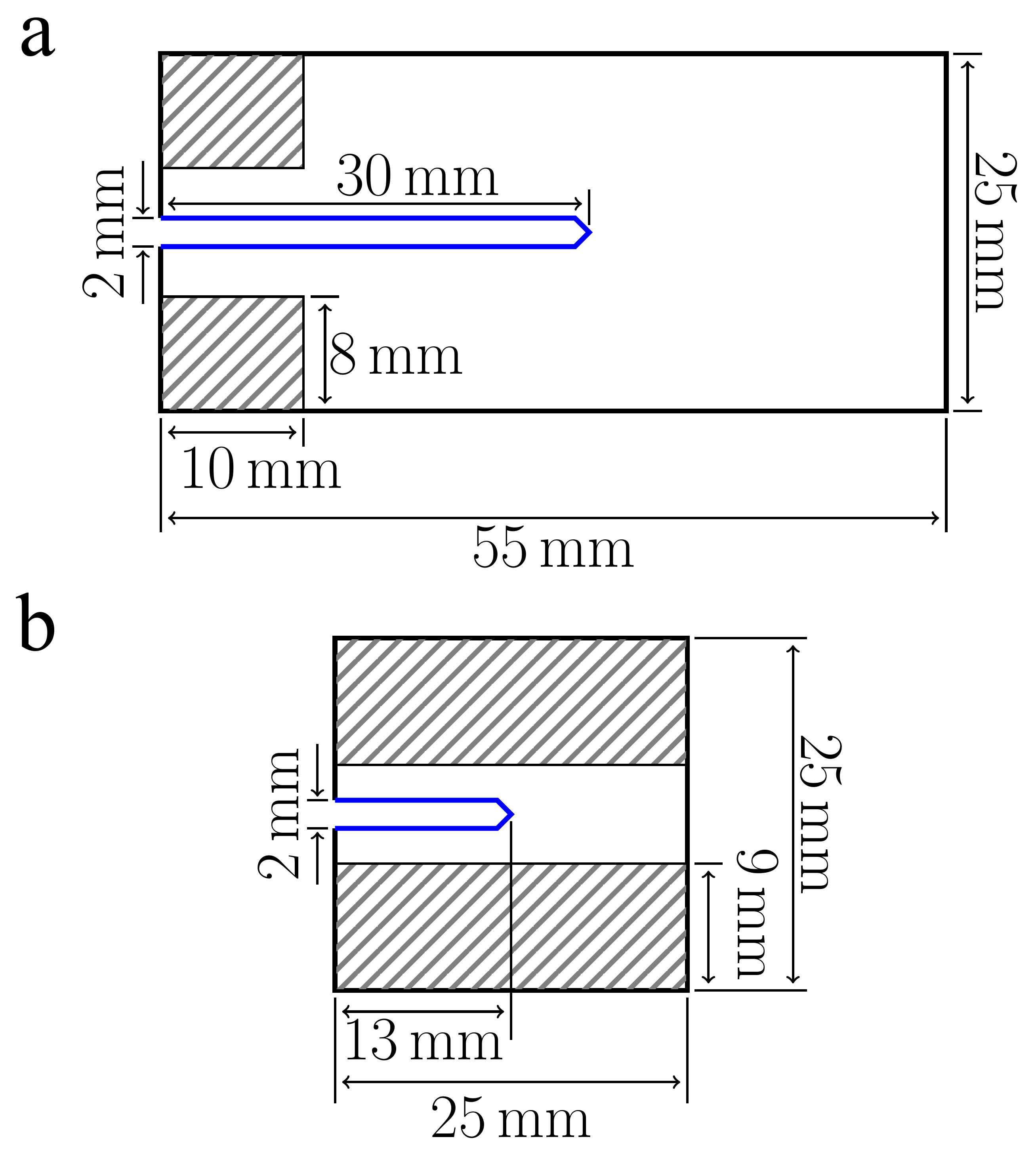}
	\caption[Schematics of experimental samples]{(a) Long and (b) Short samples. The gray areas were mounted in Instron grips.}
	\label{fig:schematics}
\end{figure}

Using our composites, we conducted Mode I tensile testing experiments of notched specimens in the two different geometries depicted in \Fig~\ref{fig:schematics} with platelets oriented at different angles ($\alpha_\Gamma$) with respect to the horizontal axis perpendicular to the tensile direction. For shorthand notation, we refer to $\alpha_\Gamma=0$ and $\alpha_\Gamma=\pi/2$ as perpendicular ($\perp$) and parallel ($\parallel$) orientations, respectively. The $\parallel$ orientation has a smaller fracture energy than the $\perp$ orientation and produces straight propagating cracks for both the short and long sample geometries. In addition, the short sample geometry suppresses kinking entirely for the $\perp$ orientation. It can therefore be used to measure experimentally the fracture energy for straight propagating cracks for both the $\parallel$ and $\perp$ orientations and to quantify the fracture energy anisotropy as described in section \ref{sec:Gc-E-estimate}. In contrast, the long sample geometry promotes kinking for the $\perp$ orientation and is used to study the effect of the magnitude of the fracture energy anisotropy on crack path selection. Mode I loading was produced by symmetrically gripping the samples at two opposite boundaries as commonly done in polymeric~\cite{kolvin2018topological} and biological~\cite{Quan:2018} materials.
The samples were mounted in tensile grips of a universal tester (Instron-5966 with $500\,\mathrm{N}$ load cell) with a data recoding frequency of $10\,\mathrm{Hz}$. The gripping regions are shown by the gray areas in \Fig~\ref{fig:schematics}.
Care was taken to ensure the grip configuration and clamping force did not play a role in the measured properties propagating cracks far from the clamps.

During the tensile loading, the bottom clamp remained fixed while the top clamp raised up at $20\,\mm/\mathrm{min}$.
For long samples, the out of plane motion was partially restrained by two glass slides separated by $1\,\mm$. 
Our experimental observations (in both short and long samples) verify that the clamping did not introduce mode-II at the crack tip and only influenced the kinking through change of the T-stress.
In both phase-field simulations and experiments, small deviations can have either small positive ($\theta^*>0$) or negative ($\theta^*<0$) angles, confirming that there is no bias. 

The crack propagation was captured by a mounted phone camera with a recording ratio of 30 frames/second. 
After the test, an image of the cracked sample was taken with a $1\,\mathrm{cm}$ grid paper underneath it and the crack shape was measured in {ImageJ}~\cite{Rueden:2017}.
The coordinates of the dots were quantified and mapped into a new coordinate system with the origin located at the notched front. We calculate the emergent angle of cracks as the linear fit to the first $2\,\mm$ of the crack path.

\subsection{Estimation of fracture toughness and elastic modulus anisotropies}\label{sec:Gc-E-estimate}

The short samples geometry and loading configuration allows us to estimate the fracture toughness from the critical imposed displacement at which sample breaks.
The short sample geometry can be thought as a $L_x\times L_y$ domain with a crack that extends from $-L_x/2$ to $l_c$ under constant displacement $u=(0,\delta)$ at $y=L_y$ and fixed at $y=0$.
Since the loading is uniform we can assume that vertical strain is almost constant in front of the crack \ie $e_{yy}=\delta/L_y$. Moreover, for a slender geometry $L_x>L_y$ we can assume $e_{xx}=0$ far from the right hand side boundary.
Therefore we can write the elastic energy per unit thickness in front of the crack as
\begin{align}
	\mathcal{B}(l_c)=&\frac{1}{2}\int_{l_c}^{Lx/2}\int_0^{L_y}\sigma_{yy}e_{yy}\,dy\,dx\nonumber\\
	&=\frac{\delta^2(\lambda+2\mu)}{2}\frac{(L_x/2-lc)}{L_y}
\end{align}
where for elastic modulus $E$ and Poisson's ratio $\nu$, $\lambda=E\nu/(1-\nu^2)$ and $\mu=E/2(1+\nu)$ are Lame constants for a thin sample (\ie in plane stress).

The energy release rate by definition is the negative change of the elastic energy due to change of the crack length (in 2D and for prescribed displacement) \ie
\begin{equation}
	G=-\left.\frac{\partial \mathcal{B}}{\partial l_c}\right|_{\delta=const}=\frac{\lambda+2\mu}{2}\frac{\delta^2}{L_y}
\end{equation}
and therefore at onset of (straight) propagation, where the load is at its maximum $f=f_{\max}$ and the imposed displacement at that instance is $\delta_c$ (see \Fig~\ref{fig:exp-measurements}a), we can write the fracture toughness as
\begin{equation}\label{eq:fav-geom-Gc}
	\Gamma=\frac{\lambda+2\mu}{2}\frac{\delta_c^2}{L_y}
\end{equation}
A similar estimate can be made by a conservation of energy argument (see~\eqref{eq:energy-conservation}) assuming that the energy stored before the onset of the propagation ($l_c=0$) is spent on the creation of new fracture surfaces \ie
\begin{equation}\label{eq:fav-geom-Gc-alt}
	\Gamma=\frac{\mathcal{B}(0)}{L_x/2}=\frac{(\lambda+2\mu)}{2}\frac{\delta_c^2}{L_y}
\end{equation}

We used \eqref{eq:fav-geom-Gc} to estimate the fracture toughness of the pure polymer matrix and 5\% $\perp$ samples from their associated load-displacement curves (see \Fig~\ref{fig:exp-measurements}a) reported in Table~\ref{tab:matprop}.
To perform the calculation (and associated error analysis), we used the elastic modulus measured in independent uniaxial measurements (Table~\ref{tab:matprop}).

To calculate the anisotropy of the elastic modulus (\ie $E_{\perp}/E_{\parallel}$) and fracture toughness (\ie $\Gamma_{\perp}/\Gamma_{\parallel}$), we used the experimentally measured load displacement curves (see \Fig~\ref{fig:exp-measurements}a for example).
We can think of our experimental specimens as complex springs where their stiffness $S$ only depends on the sample geometry, loading configuration, and elastic properties. 
When the crack propagates, the resultant force drops since the sample becomes more compliant.
Therefore, at the onset of fracture, the load is at its maximum $f=f_{\max}$ where we signify the imposed displacement at that instant as $\delta_c$.
To calculate the stiffness of the sample, we fit the load per unit thickness $f/h$ of the sample (where $h$ is the sample thickness) vs. its imposed boundary displacement $\delta$ for $\delta<0.1\,\mm$ linearly.
This limit was chosen such that it is always below the critical imposed displacement $\delta_c$ for all performed experimental measurements. 
\ie the load per unit thickness $h$ is a linear function of the imposed displacement before the cracks propagate when $\delta<\delta_c$
\begin{equation}\label{eq:energy-conservation}
	\frac{f}{h}=S\delta\quad\delta<\delta_c
\end{equation}
The stiffness of the samples changes with changing $f_v$ but, for given $f_v$, the slopes of load displacement curves for the $\perp$ and $\parallel$ samples remain almost constant, which allows us to treat these samples in both $\perp$ and $\parallel$ directions as isotropic.

Next we show that the fracture toughness anisotropy can be calculated from the work to fracture. 
Due to energy conservation, we can write the energy stored per unit thickness $\mathcal{B}$ in terms of the imposed displacement $\delta$ and the resultant force $f$ in short samples (before the cracks advance) as
\begin{equation}
	\mathcal{B}(\delta)=\int_{\Omega}\mathcal{W}(\ub)\,d\xb=\frac{1}{2}\int_0^{\delta} \left(\frac{f}{h}\right)\,d\delta
\end{equation}
here $\mathcal{W}(\ub)=(C_{ijkl}e_{ij}e_{kl})/2$ is the elastic energy density where $C_{ijkl}=\lambda\delta_{ij}\delta_{kl}+\mu(\delta_{il}\delta_{jk}+\delta_{ik}\delta_{jl})$ is the linear elastic constitutive tensor and $e_{ij}=(u_{i,j}+u_{j,i})/2$ is the linear strain.

Since in ideally brittle materials fracture is the only dissipating mechanism, we can estimate the fracture toughness by matching the energy $h\mathcal{B}(\delta_c)$ released when the crack breaks the sample in two to the energy $\Gamma hl_c$ needed to create the crack of length $l_c$, which yields 
\begin{equation}
	\mathcal{B}(\delta_c)=\Gamma l_c
\end{equation}
Therefore, we can write the ratio of the $\perp$ and $\parallel$ fracture toughnesses as
\begin{equation}\label{eq:Gc-anisotropy}
	\frac{\Gamma_{\perp}}{\Gamma_{\parallel}}=\frac{\gamma_{f\perp}}{\gamma_{f_{\parallel}}}
\end{equation}
where the different subscripts correspond to different orientations of platelets and we define the work to fracture per unit thickness as $\gamma_f$ as the area of load per unit thickness vs. displacement curve
\begin{equation}\label{eq:fracture-work}
	\gamma_f=\int_0^{\delta_c}\left(\frac{f}{h}\right)\,d\delta
\end{equation}
Thus, the straight crack propagation for both $\perp$ and $\parallel$ in the short samples (see \Fig~\ref{fig:crack-path-exp}) enables us to accurately measure the anisotropy of fracture toughness from the measured load-displacement curves (\eg \Fig~\ref{fig:exp-measurements}a).


\subsection{Phase-field modeling}

The phase-field model couples the elastic displacement field $\ub=(u_x,u_y)$ to a scalar phase field $\phi$ that varies smoothly from $\phi=1$ in the pristine material to $\phi=0$ in the fully broken material over a length scale $\xi$, which sets the size of the process zone around the crack tip where linear elasticity breaks down. 
The total energy of the system is described by the functional 
\begin{align}
	\F_{\xi}(\ub,\phi)=&\int_{\Omega}g(\phi)\mathcal{W}(e(\ub))\,d\xb \nonumber\\&+\frac{\Gamma_{\perp}}{4C_{\phi}}\int_{\Omega}\left(\frac{w(\phi)}{\xi}+\xi A_{ij}\partial_{x_{i}x_{j}}\phi\right)\,d\xb
	\label{eq:ATE}
\end{align}
where the first and second terms on the right-hand-side correspond to the elastic strain energy and the anisotropic fracture energy~\cite{Hakim:2009}, respectively.
We define the fracture energy anisotropy matrix as
\begin{align}
	\hat{A}&=\begin{bmatrix}
    			\A^{-2} & 0 \\ 0 & 1
				\end{bmatrix}\label{eq:A-matrix}\\
	A_{ij}&=Q_{ik}\left(\frac{\pi}{2}-\alpha_\Gamma\right)Q_{jl}\left(\frac{\pi}{2}-\alpha_\Gamma\right)\hat{A}_{kl}\label{eq:alpha-rotation}
\end{align}
where $Q$ is the rotation matrix, and we denote by $\alpha_\Gamma$ the angle between the plane of the platelet and the horizontal axis such that $\alpha_\Gamma=0^\circ$ and $\alpha_\Gamma=90^\circ$ correspond to the $\parallel$ and $\perp$ orientations, respectively.
With the choice $C_{\phi}=\int_{0}^{1}\sqrt{w(\phi)}\,d\phi$, and for $\perp$ orientations (\ie $\alpha_\Gamma=\pi/2$) the fracture energy is $\Gamma_\perp$ for propagation along the $x$ direction parallel to the parent crack and perpendicular to the platelet faces and $\Gamma_\parallel=\Gamma_\perp/\A$ for propagation along $y$ parallel to the faces. 
In particular, for $\perp$ orientations, Eqs.~\eqref{eq:ATE}--\eqref{eq:alpha-rotation} results in an anisotropic fracture energy of the form~\cite{Hakim:2009}
\begin{align}\label{eq:anisotropy}
	\Gamma(\theta)=\Gamma_{\perp}\sqrt{\A^{-2}\sin^2(\theta)+\cos^2(\theta)}
\end{align}
Furthermore, $\mathcal{W}(e(\ub))$ is the elastic energy density defined for isotropic linear elastic solid as $\mathcal{W}(e(\ub))=(\mathcal{C}_{ijkl}\,e_{kl}({\ub})e_{ij}({\ub}))/2$ where $e_{ij}(\ub)=(\partial_{x_{j}} u_{i} +\partial_{x_{i}} u_{j})/2$ is the strain tensor and the elastic constitutive tensor for plane-stress elasticity is given as $\mathcal{C}_{ijkl}=\lambda\delta_{ij}\delta_{kl}+\mu(\delta_{il}\delta_{jk}+\delta_{ik}\delta_{jl})$ where
$\lambda=E\nu/(1-\nu^2)$ and $\mu=E/(2(1+\nu))$ are the Lame coefficients for elastic modulus $E$ and Poisson's ratio $\nu$.

In addition, we use the specific forms of the function $g(\phi)=4\phi^3-3\phi^4$ and $w(\phi)=1-g(\phi)$~\cite{Karma:2001a,Hakim:2009} to model the propagation of a fracture from a single flaw by prohibiting the initiation of new cracks in the undamaged ($\phi=1$) material. 
All simulations are performed in 2D plane stress using the classic iterative minimization for quasi-static crack propagation~\cite{Bourdin:2008a}, which consists of finding the minimizers of $\F_{\xi}$ by solving the Euler-Lagrange equations for~\eqref{eq:ATE}. 
The Euler-Lagrange equations derived variationally from Eq.~\eqref{eq:ATE} are discretized using the Galerkin finite element method and solved using distributed data structures provided by {libMesh}~\cite{libmesh} and linear algebra solvers in {PETSc}~\cite{petsc-efficient,petsc-user-ref}. 
The sample geometries depicted in \Fig~\ref{fig:schematics} are meshed using a triangular Delaunay mesh with average edge size $\approx 27.5\,\micron$.
To perform the numerical simulations, we imposed the boundary conditions associated with the grips as $u_x=0,u_y=\pm\delta$ on all nodes in contact with the grips (\ie the gray shaded areas in \Fig~\ref{fig:schematics} and \Fig~\ref{fig:exp-measurements}{a}) and the sharp notch was simulated by imposing $\phi=0$ at the tip of the v-shaped notch.
Typical simulation included $\approx500\,\mathrm{kDOFs}$ and was ran on 40 physical cores of 2.2 GHz Intel Xeon E5-2630 CPU for $\approx\,24\mathrm{hr}$. 
Simulations are carried out with the estimate $\nu=0.2$, and the input parameters $\A$ and $\xi$. 

Finally, we calculate the initial kink angle $\theta^*$ by measuring the angle of the line that connects the crack tip to the initial notch tip at the first time step where the crack is propagated a distance larger than $2\xi$. 
The standard deviation of measurement is calculated as the maximum change in angle as the result of discretization. 

\subsection{Dimensional analysis of the phase-field model}\label{sec:PFM-dimensional-analysis}

Since accurate values of the fracture toughnesses $\Gamma_{\perp}$ and $\Gamma_{\parallel}$ cannot be calculated directly from the experimental measurements, we show in this section that the crack path is only affected by the ratio of these energies (the fracture toughness anisotropy $\A=\Gamma_{\perp}/\Gamma_{\parallel}$) and not their individual values.
If we define the dimensionless coordinates $\bar{\xb}=\xb/L$ (where we choose $L$ to be the sample width), dimensionless displacement $\bar{\ub}=\ub/\sqrt{\Gamma_{\perp}L/E}$, and dimensionless elasticity tensor $\bar{\C}=\C/E$, using Eq.~\eqref{eq:ATE} we can write the dimensionless energy as
\begin{align}\label{eq:ATE-nD}
	\bar{\F}_{\xi}(\bar{\ub},\phi)&=\frac{{\F}_{\xi}(\bar{\ub},\phi)}{\Gamma_{\perp}}=\int_{\Omega}g(\phi)\bar{\mathcal{W}}(e(\bar{\ub}))\,d\bar{\xb}\nonumber\\+&\frac{1}{4C_{\phi}}\int_{\Omega}\left(\frac{w(\phi)}{\bar{\xi}}+\bar{\xi}A_{ij}\partial_{\bar{x}_{i}\bar{x}_{j}}\phi\right)\,d\bar{\xb}
\end{align}
where $\bar{\mathcal{W}}(e(\bar{\ub}))=\bar{\mathcal{C}}_{ijkl}\bar{e}_{kl}(\bar{\ub})\bar{e}_{ij}(\bar{\ub})/2$, $\bar{e}_{ij}(\bar{\ub})=(\partial_{\bar{x}_{j}} \bar{u}_{i} +\partial_{\bar{x}_{i}} \bar{u}_{j})/2$, and $\bar{\xi}=\xi/L$. 
Therefore it is easy to see that the crack path predicted by this model only depends on the sample geometry and loading configuration, the relative size of process zone with respect to the sample size $\xi/L$, Poisson's ratio $\nu$, and the fracture energy anisotropy $\A=\Gamma_{\perp}/\Gamma_{\parallel}$.

\subsection{Determination of the T-stress in short and long samples}\label{ssec:T-stress-approx}
{To assess the role of the T-stress in crack kinking we have to calculate the T-stress in long and short samples. 
The results of these computation are later used in section~\ref{ssec:sample-geometry-effect} for quantitative comparison of analytical predictions of crack kinking behavior with experiments and phase-field simulations.
For a crack along the x-axis in an isotropic homogeneous elastic media the near crack tip stress fields for plane-stress can be written as}
\begin{align}
\sigma_{xx}(r,\theta)&=\frac{K_I}{\sqrt{2\pi r}}\cos\left(\frac{\theta}{2}\right)\left[1-\sin\left(\frac{\theta}{2}\right)\sin\left(\frac{3\theta}{2}\right)\right]\nonumber\\&-\frac{K_{II}}{\sqrt{2\pi r}}\sin\left(\frac{\theta}{2}\right)\left[2+\cos\left(\frac{\theta}{2}\right)\cos\left(\frac{3\theta}{2}\right)\right]\nonumber\\&+T+\mathcal{O}(\sqrt{r})\label{eq:crack-tip-sxx}
\end{align}
\begin{align}
\sigma_{yy}(r,\theta)&=\frac{K_I}{\sqrt{2\pi r}}\cos\left(\frac{\theta}{2}\right)\left[1+\sin\left(\frac{\theta}{2}\right)\sin\left(\frac{3\theta}{2}\right)\right]\nonumber\\&+\frac{K_{II}}{\sqrt{2\pi r}}\sin\left(\frac{\theta}{2}\right)\cos\left(\frac{\theta}{2}\right)\cos\left(\frac{3\theta}{2}\right)\nonumber\\&+\mathcal{O}(\sqrt{r})\label{eq:crack-tip-syy}
\end{align}
where $K_I,K_{II}$ are the stress-intensity-factors ({SIF}s), $T$ is T-stress, and $(r,\theta)$ are the polar coordinates centered at the crack tip.
Using \eqref{eq:crack-tip-sxx}-\eqref{eq:crack-tip-syy}, it is easy to see that the T-stress can be calculated from the divergent stress fields at the crack tip as  
\begin{equation}\label{eq:sT-estimate}
	T={\sigma}_{xx}(r,0)-{\sigma}_{yy}(r,0)
\end{equation}

\begin{figure}[ht!]
	\centering
	\includegraphics[width=.45\textwidth]{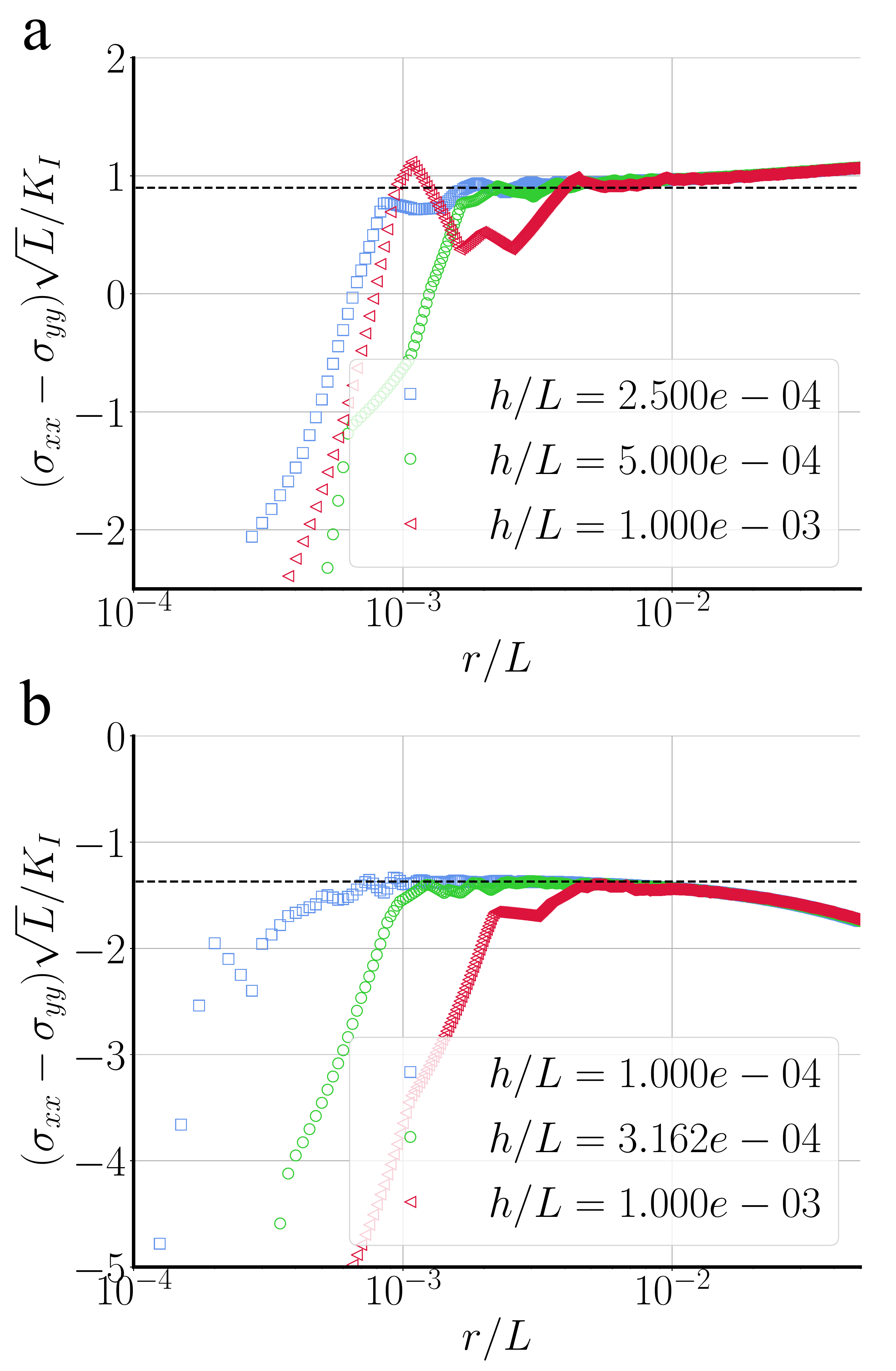}
	\caption[T-stress calculation in short and long samples]{Results of numerical simulations for (a) the long sample ($L=55\,\mm$) showing convergence for calculated value of T-stress $T\sqrt{L}/K_I\approx0.9$ for three different mesh size $h/L=2.5\times10^{-4},5\times10^{-5},10^{-3}$ (b) the small sample ($L=25\,\mm$) showing convergence for calculated value of T-stress $T\sqrt{L}/K_I\approx-1.37$ for three different mesh size $h/L=1\times10^{-4},3.162\times10^{-5},10^{-3}$.}
	\label{fig:sT-long-short}
\end{figure}

{To accurately estimate the T-stress value and mimic the razor blade notch in the experiments, we replaced the wide pre-notches in the two samples (blue lines in \Fig~\ref{fig:schematics}) with a sharp notch with opening angle $1^{\circ}$ and refined the mesh around the crack tip to capture the singularity. 
To estimate the T-stress in the two samples, we numerically calculate $K_I$ for given imposed vertical displacement $\delta$.
The mode-I {SIF} $K_I=\sqrt{GE}$ (for plane-stress) was calculated by first calculating the energy release rate $G$ using the G-$\theta$ method~\cite{Destuynder:1981}, which replaces the J integral by a surface integral (a volume integral in 3-D) using Stokes' theorem.
\Fig~\ref{fig:sT-long-short} shows the results of the numerical simulation where we plotted $({\sigma}_{xx}(r,0)-{\sigma}_{yy}(r,0))\sqrt{L}/K_I$ against the dimensionless distance from the crack tip $r/L$ near the crack tip ($L=55\,\mm$ in the long geometry and $L=25\,\mm$ in the short geometry see \Fig~\ref{fig:schematics}). 
It should be noted that, since the simulations are performed in a finite domain, the divergent stress forms are only accurate in the vicinity of the crack tip.
Moreover, since we are using $C_0$ continuous elements with no enrichment, the stress fields are not accurate at very short distances near the tip $r/l<\mathcal{O}(h/L)$ where $h$ is the mesh size.}

\section{Results}\label{sec:results}
\subsection{Experimental results for \texorpdfstring{$\parallel$}{parallel} and \texorpdfstring{$\perp$}{perpendicular} platelet orientations}

\begin{figure*}[htb]
\begin{center}
\includegraphics[width=1.8\columnwidth]{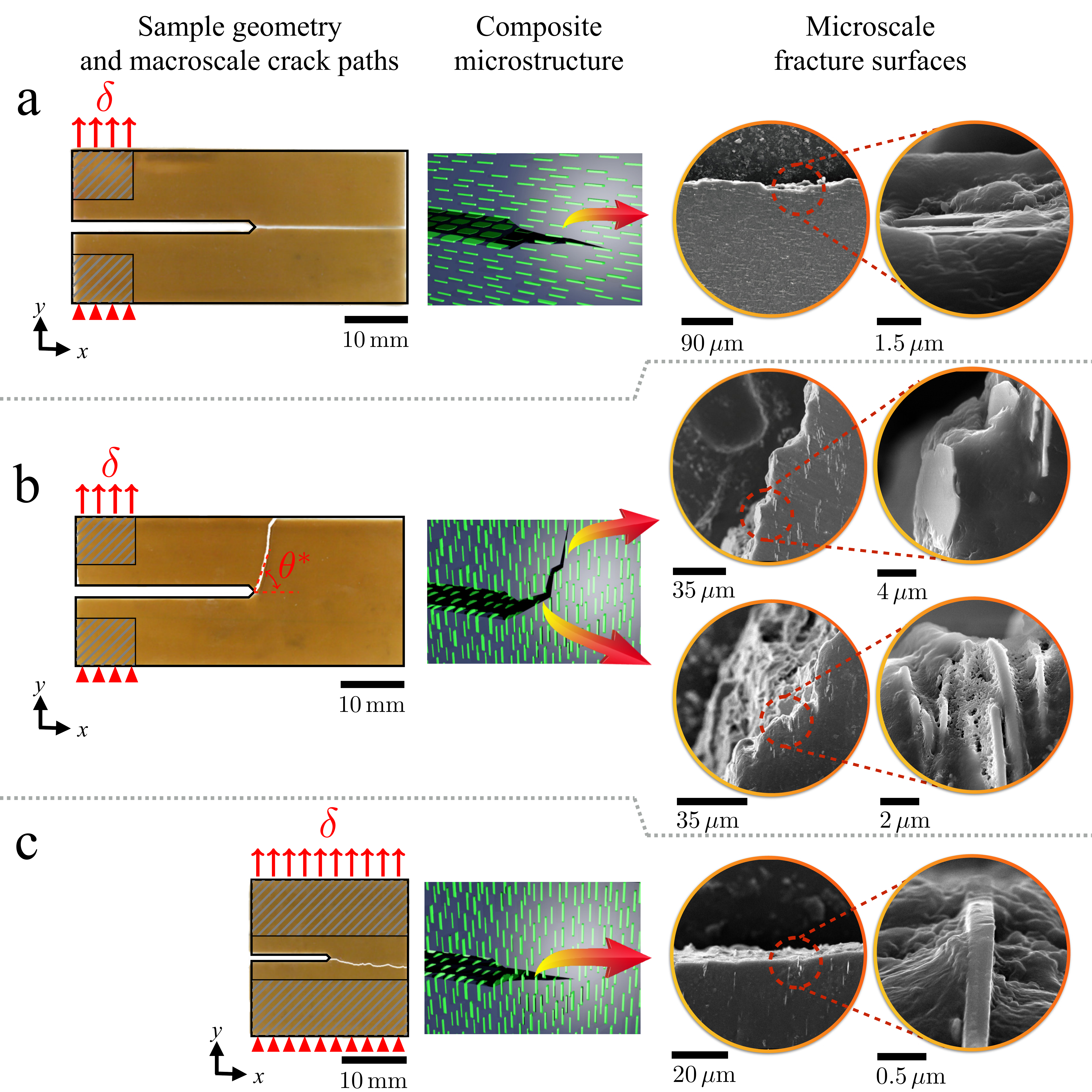}
\caption{
Results of pure tensile fracture experiments for 5\% platelet volume fraction bicomposites illustrating stark differences between microscale and macroscale crack paths and the strong influence of sample geometry on macroscale paths. The left column shows the macroscale crack paths in fractured samples for different platelet orientations illustrated in the middle column and the right column shows SEM images of microscale crack paths at different magnification.  
\textbf{a}, Long sample with platelets oriented parallel to the crack propagation axis. The macroscale crack path is straight and the microscale fracture path travels along the platelet faces. The same behavior is observed in short samples (results not shown). 
\textbf{b}, Long sample with platelets oriented perpendicular to the crack propagation axis. The macroscale crack path is strongly kinked and the microscale fracture path travels along a staircase with vertical sections parallel to the platelet faces (top two images or right column) and horizontal sections perpendicular to the faces (bottom two images of right column). \textbf{c}, Short sample with the same platelet orientation as in \textbf{b}. In contrast to \textbf{b}, the macroscale crack path remains straight even though the platelets deflect the crack on a microscale. Platelets do not break and deflect cracks on a microscale in all samples.}
\label{fig:figure1}
\end{center}
\end{figure*}

\begin{figure*}[htb]
\begin{center}
\includegraphics[width=2\columnwidth]{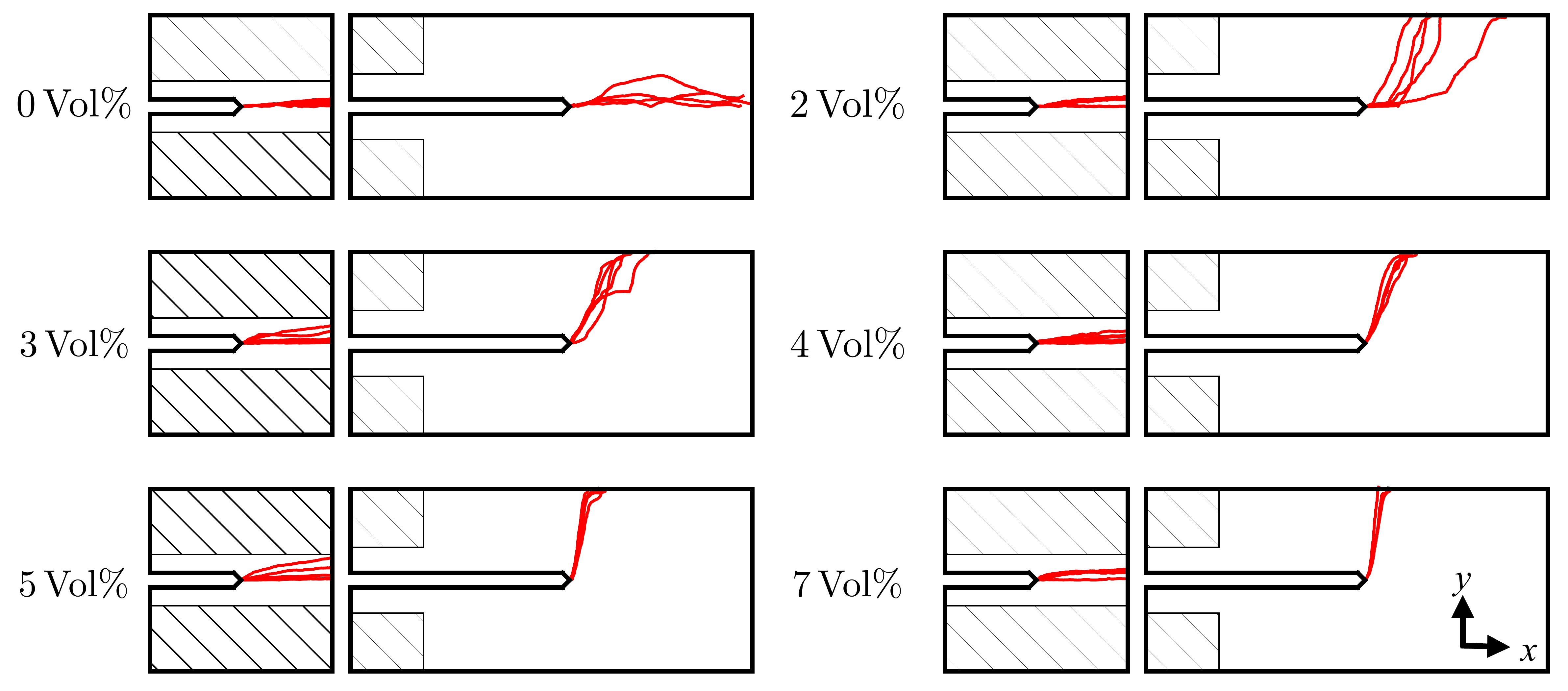}
\caption{
Experimentally observed macroscale crack paths in short and long samples for different volume fractions of alumina platelets oriented perpendicularly to the crack propagation axis ($\perp$ orientation). The crack paths are digitized 
and \underline{mirrored to $y>0$} to show consistency of the initial crack kinking angle $\theta^*$. Crack paths remain straight in all short samples but kink in long samples above a critical volume fraction of approximately 3\%.
}
\label{fig:crack-path-exp}
\end{center}
\end{figure*}

\begin{figure}[htb!]
\begin{center}
\includegraphics[width=\columnwidth]{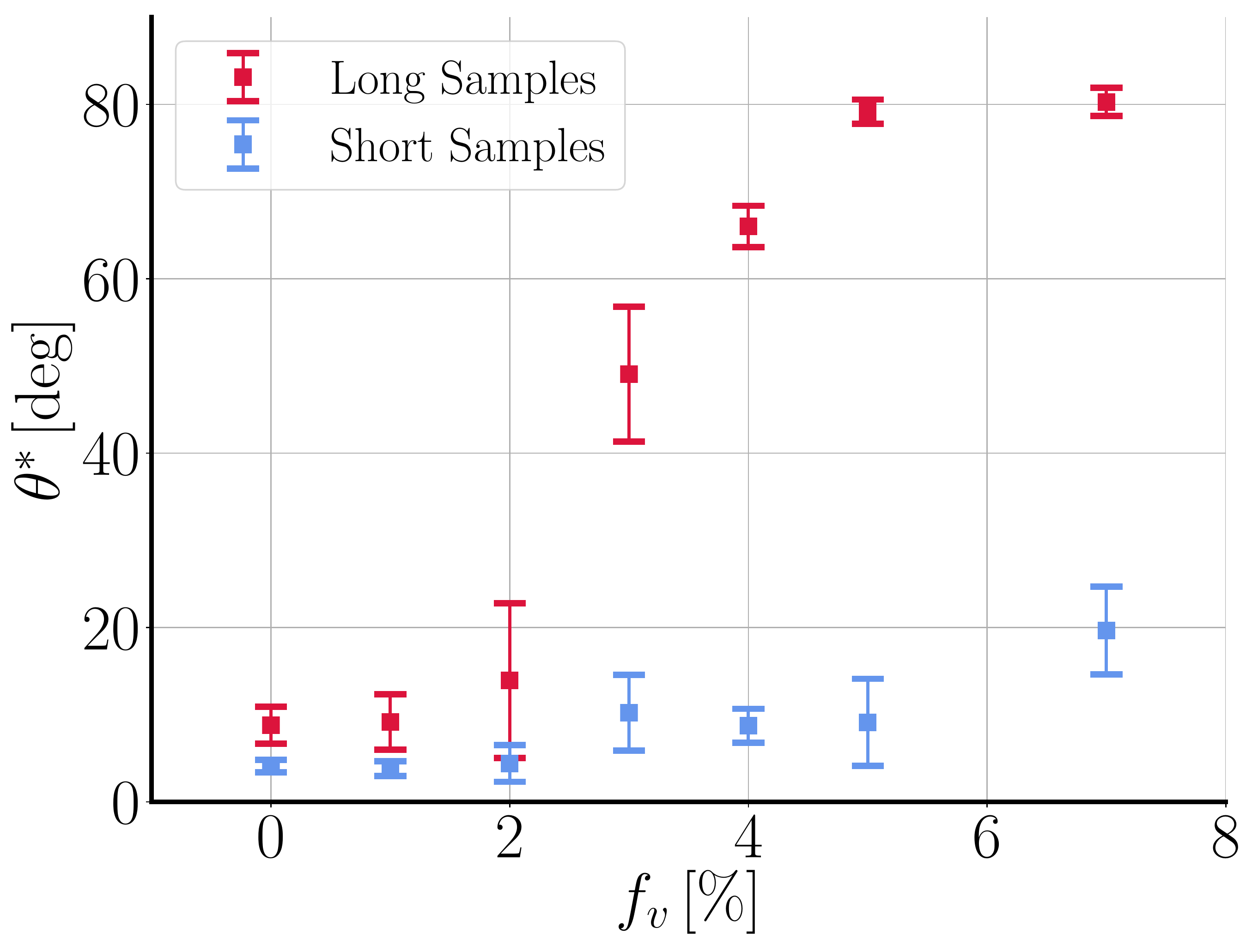}
\caption{
Measured crack kinking angle $\theta^*$ from the experimental results in \Fig~\ref{fig:crack-path-exp}. 
}
\label{fig:angle-exp}
\end{center}
\end{figure}

\Fig~\ref{fig:figure1} shows the results of Mode I fracture experiments in different geometries with $\parallel$ and $\perp$ platelet orientations 
and volume fraction of platelets $f_v$ from $0$ to $0.07$.
In the long sample geometry, crack kinking occurs with perpendicular oriented platelets for sufficient volume fraction, while in the short sample cracks propagate straight across for all volume fractions in this range. Examination of crack paths on a microscale (\Fig~\ref{fig:figure1} right columns) revealed that the crack front did not penetrate the platelets that are orders of magnitude stronger and stiffer than the matrix as shown in Table \ref{tab:matprop}. 
As a result, the crack front propagated around platelets following a tortuous microscale path. 
Examination of crack paths on a macroscale (shown in \Fig~\ref{fig:crack-path-exp} and quantified in \Fig~\ref{fig:angle-exp}) revealed that, for the $\perp$ orientation, cracks propagated straight in all short samples over the range of volume fraction $f_v\leq0.07$ (\Fig~\ref{fig:angle-exp}), despite being sporadically deflected on a microscale (see \Fig~\ref{fig:figure1}c right column).
In contrast, in long samples cracks exhibited a clear transition from straight to kinked propagation over the same range of $f_v$. 
Several experiments were conducted in both sample geometries to show that crack paths were highly reproducible.
Those observations demonstrate that microscale crack deflection, common in composites with hard particles embedded in a softer matrix, is not generally a sufficient condition for macroscale deflection, which depends in a non-trivial way on both the microstructure ($f_v$) and sample/loading geometry.       

\begin{figure*}[!htb]
\begin{center}
\includegraphics[width=2\columnwidth]{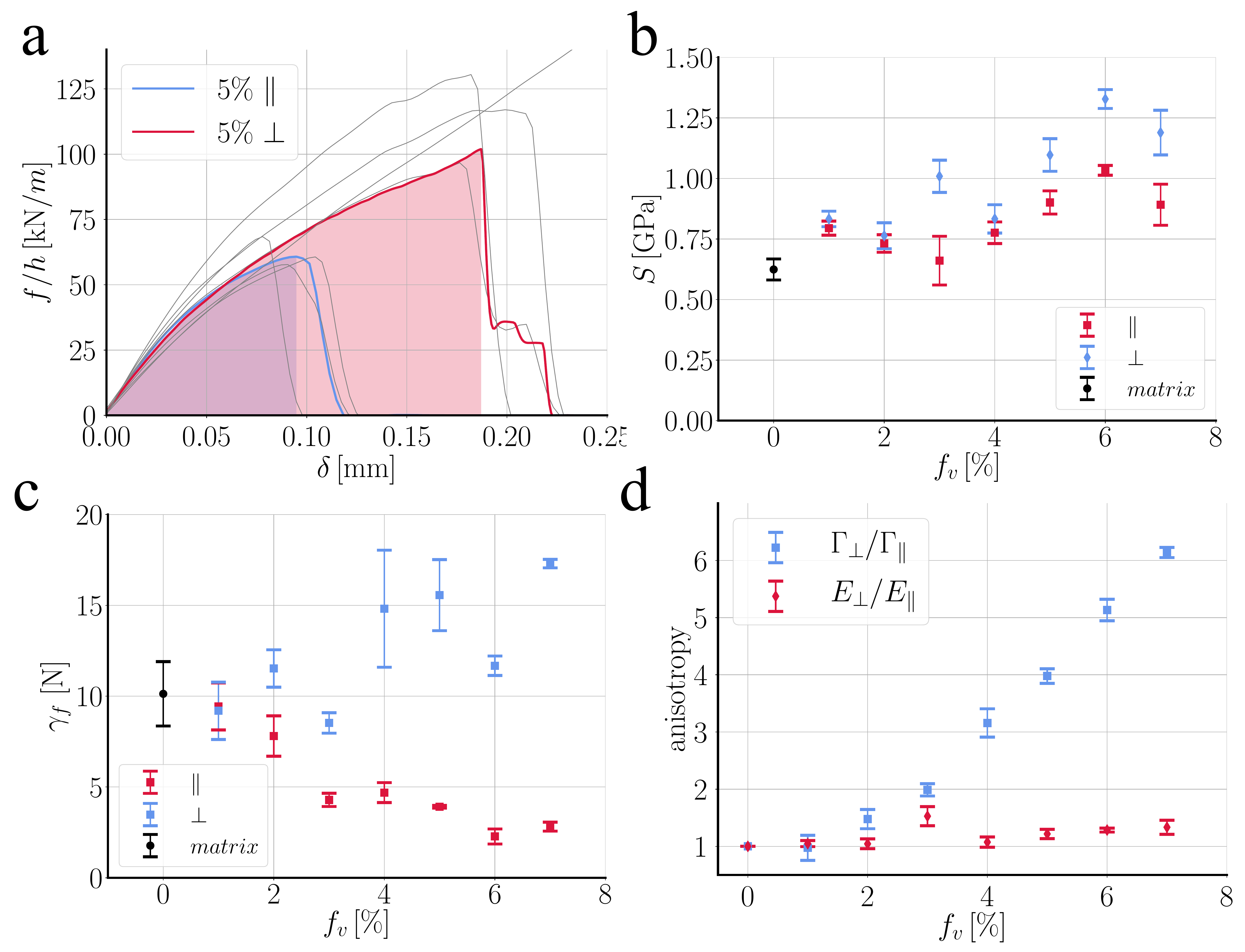}
\caption{
\textbf{a}, Illustration of load-displacement curves used to measure the fracture energy anisotropy by exploiting the fact that cracks propagate straight in short samples for both $\parallel$ and $\perp$ platelet orientations. Curves for several $5\,\vol$ samples are shown as gray lines. Fracture occurs at higher load and displacement for the $\perp$ orientation, reflecting a higher fracture energy. The colored lines and shaded regions illustrate the area under the curve used for the calculation of the fracture energy anisotropy for one $\parallel$ sample (blue) and one $\perp$ sample (red).
\textbf{b}, Calculated stiffness of short samples for sample made of pure matrix as well as samples with $\parallel$ and $\perp$ platelet orientations.
\textbf{c}, Estimated fracture energy from short samples for pure matrix as well as $\parallel$ and $\perp$ platelet orientations.
\textbf{d}, Fracture energy anisotropy $\A=\Gamma_{\perp}/\Gamma_{\parallel}$ (blue) and elastic modulus anisotropy $E_{\perp}/E_{\parallel}$ (red) along with their standard errors versus volumetric percentage of alumina platelets $f_v$ over the whole range of $f_v$ where cracks remain straight in short samples for both $\parallel$ and $\perp$ orientations.
}
\label{fig:exp-measurements}
\end{center}
\end{figure*}

As a first step towards crack path prediction on a macroscale, we recorded load-displacement curves (\Fig~\ref{fig:exp-measurements}a) to measure mechanical properties of composites relevant for fracture, including the anisotropies of the fracture energy and elastic modulus respectively defined as the ratios $\Gamma_{\perp}/\Gamma_{\parallel}$ and $E_{\perp}/E_{\parallel}$ of those quantities for the $\perp$ and $\parallel$ orientations. 
We used short samples in which cracks propagate straight on the macroscale, thereby allowing us to perform a direct measurement of $\Gamma_{\perp}/\Gamma_{\parallel}$ for all volume fractions, which is not feasible in long samples that exhibit kinking.
$E_{\perp}/E_{\parallel}$ was computed as the ratio of the slopes of the load-displacement curves for small displacements and $\Gamma_{\perp}/\Gamma_{\parallel}$ as the ratio of the areas under those curves up to fracture initiation.
We note that the work to fracture, which is commonly used as an approximate measure of toughness~\cite{mirkhalaf2014overcoming,kamat2000structural,Shin:2012},  was not used here to measure the fracture energy for a given orientation, but only the ratio $\Gamma_{\perp}/\Gamma_{\parallel}$ for the $\perp$ and $\parallel$ orientations. 
Since the same short sample geometry is used to compute the areas under the force-extension curves for those two platelet orientations, and crack propagate straight in this geometry for $f_v<0.07$ (thereby producing the same fracture surface area), the ratio of work to fracture is equal to the ratio of fracture energies for those two orientations as discussed in section~\ref{sec:Gc-E-estimate}. 
While those slopes measure the stiffnesses of the samples and generally depend on sample geometries, the ratio of stiffnesses for the $\perp$ and $\parallel$ orientations is identical to the ratio $E_{\perp}/E_{\parallel}$ since identical geometries are used for both orientations.
Results in \Fig~\ref{fig:exp-measurements}d (\Fig~\ref{fig:exp-measurements}(b-c) for individual measurements) show that, when $f_v$ increases from 0 to 7\%, $\Gamma_{\perp}/\Gamma_{\parallel}$ increases about 600\% while $E_{\perp}/E_{\parallel}$ only increases by about 20\%.
This suggests that the fracture energy anisotropy is predominantly responsible for crack kinking in long samples, but leaves open the question of why it is absent in short samples over the same range of $f_v$.  

\subsection{Phase-field modeling results for \texorpdfstring{$\parallel$}{parallel} and \texorpdfstring{$\perp$}{perpendicular} platelet orientations}
To gain more insight into these results, we use the phase-field method~\cite{Bourdin:2008a,Karma:2001a,Hakim:2009} described in Section~\ref{sec:methods} to model crack paths on a macroscale using the experimentally measured fracture energy anisotropy and a quantitative estimate of the process zone size as input into the model.  
To estimate the process zone size, we performed independent measurements of material properties of the polymer matrix and 5\% $\perp$ composite using simple uniaxial tension. 
We chose the tensile axis parallel to the platelets corresponding to the $\perp$ orientation in our fracture experiments.
The elastic modulus $E$, tensile strength $\sigma_{c}$ where measured using ASTM-D638V uniaxial tension test.
In addition to above uniaxial tests, the fracture energy $\Gamma_\perp$ of the matrix and the 5\% composite was estimated based on value of imposed displacement at the onset of fracture $\delta_c$ in the experimental load-displacement curves of the short samples (see Section~\ref{sec:Gc-E-estimate}).
Table~\ref{tab:matprop} summarizes the measured elastic modulus $E$, tensile strength $\sigma_{c}$, and the fracture energy of different composite components and the resulting composite.
We further use the theoretical estimate $\xi\sim\Gamma_{\perp}E_\perp/\sigma_{c\perp}^2$ that follows from assuming that the maximum opening stress $\sigma_{yy}\sim K_I/\sqrt{\xi}$ at the crack tip is comparable to $\sigma_{c\perp}$. 

\begin{table*}[htb!]
\begin{center}
\caption{Independently measured elastic modulus $E$ and maximum tensile strength $\sigma_{c}$, and calculated fracture energy estimate (see~\ref{sec:Gc-E-estimate}) along with their associated standard errors for the polymer matrix, and the 5\% composite. The corresponding values reported for the Alumina platelets from~\cite{Auerkari:1996}.}\label{tab:matprop}
\vspace{.5em}
\begin{tabular}{l c c c}
\toprule
Material & $E\,[\mathrm{GPa}]$&$\sigma_{c}\,[\mathrm{MPa}]$&$\Gamma\,[\mathrm{kJ/m^2}]$ \tabularnewline
\midrule
Alumina platelets~\cite{Auerkari:1996}	& $380\textup{--}410$	& $210\textup{--}500^\dagger$	& $0.022\textup{--}0.095^{\ddagger}$ \tabularnewline
Polymer Matrix 							& $0.444\pm0.011$		& $22.53\pm0.78$				& $0.84\pm0.18$\tabularnewline
5\% $\perp$ Composite 				& $0.603\pm0.021$ 	 	& $25.22\pm0.97$ 				& $1.6\pm0.5$\tabularnewline
5\% $\parallel$ Composite 			& $0.472\pm0.08$ 	 	& $14.16\pm1.3$ 				& $0.34\pm0.06$\tabularnewline
\bottomrule
\multicolumn{2}{c}{\small$\dagger$ Flexural strength. $\ddagger$ Estimated from $K_{IC}$} & \multicolumn{2}{c}{} \tabularnewline
\end{tabular}
\end{center}
\end{table*}

The values of $E$ and $\Gamma$ only determine the physical magnitude of the imposed displacement $\delta\sim \sqrt{\Gamma_L/E}$ that produces the applied tensile load, where $L$ is the sample width, but do not affect the fracture behavior. 
The process zone size $\xi$ is estimated by assuming that the maximum opening stress on the process zone scale $\sigma_{yy}\sim K_I/\sqrt{\xi}$ is comparable to the tensile strength of the material $\sigma_c$. 
Setting $K_I$ equal to its value $K_{IC\perp}=\sqrt{\Gamma E}$ at the onset of propagation of a straight crack in short samples, yields the estimate $\xi\sim\Gamma E/\sigma_c^2$.
By performing phase-field simulations we obtain the proportionality factor as 
\begin{equation}
	\xi\simeq0.39\,\dfrac{\Gamma E}{\sigma_{c\perp}^2}
	\label{eq:process-zone}
\end{equation}
A similar estimate has been previously obtained~\cite{Pham:2012,Pham:2013} by a one-dimensional stability analysis of Eq.~\eqref{eq:ATE} for a broad class of functions $g(\phi)$ and $w(\phi)$ that did not include the present model.
Combining Eq.~\eqref{eq:process-zone} with our measurements of the elastic modulus and the ultimate tensile strength along with our estimate of fracture energy (compiled in Table~\ref{tab:matprop}), we estimate the process zone size as $\xi\simeq285\pm126\,\micron$ for the polymer matrix, $\xi\simeq309\pm115\,\micron$ for the 5\% $\parallel$ composite, and $\xi\simeq593\pm207\,\micron$ for the 5\% $\perp$ composite (where $\pm$ signs signify standard errors).
For the matrix, this $\xi$ estimate is comparable to the length of a craze region. For the composite, it is much larger than the mean platelet spacing ($\simeq\,10\micron$) consistent with previous estimates that $\xi$ is approximately $5\textup{--}50$ times larger than the microstructure scale in diverse composites~\cite{Bazant:2004}.
We should also highlight that reduction in the process zone size $\xi$ from $\perp$ to $\parallel$ roughly follows the prediction from our phase-field model \ie $\xi_\perp/\xi_\parallel=\A$. 
We use for all the computations $\xi=225$ and $550\,\micron$ as the lower and upper bounds of process zone size, respectively.

\begin{figure}[!htb]
\begin{center}
\includegraphics[width=\columnwidth]{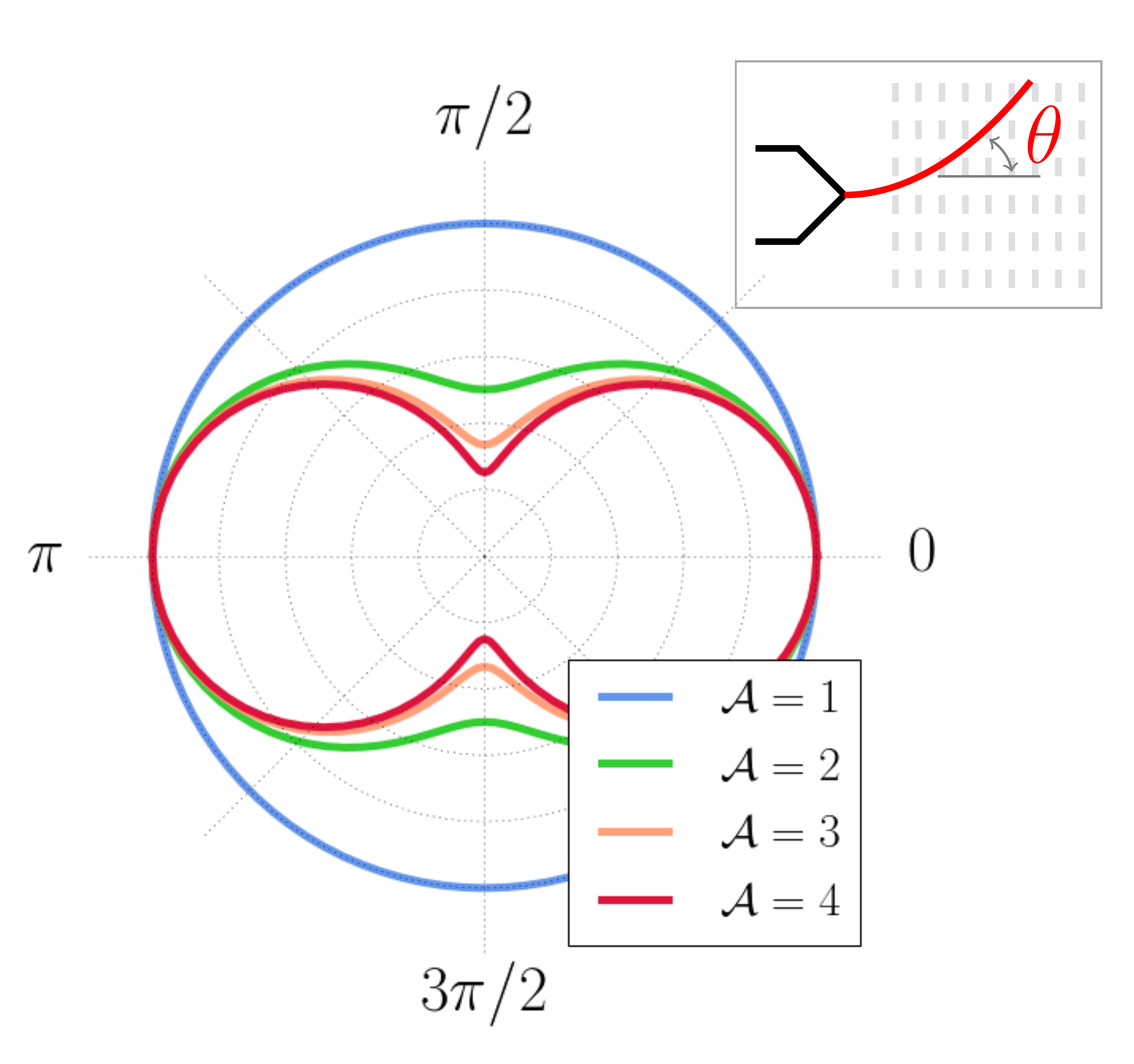}
\caption{
Polar plots of the fracture energy $\Gamma(\theta)=\Gamma_{\perp}\sqrt{\A^{-2}\sin^2(\theta)+\cos^2(\theta)}$~\cite{Hakim:2009} for different anisotropies $\A$ and $\theta$ defined as the angle between the crack propagation axis and a horizontal axis parallel to the parent crack (inset). Plots correspond to the $\perp$ orientation where platelet faces are perpendicular to the parent crack axis.
}
\label{fig:anisotropy-function}
\end{center}
\end{figure}

Since experimental samples are thin, we model fracture in 2D plane stress and focus on the crack-kinking $\perp$ orientation. 
Moreover, since $E_{\perp}/E_{\parallel}$ is weakly dependent on $f_v$, we assume that elasticity is isotropic and model the anisotropy of the fracture energy with the simple form in Eq.~\eqref{eq:anisotropy} where $\theta$ (\Fig~\ref{fig:anisotropy-function} inlay) is the angle between the crack axis and the reference straight propagation axis and $\A\equiv \Gamma_{\perp}/\Gamma_{\parallel}$ is the fracture energy anisotropy. 
This form is consistent with a 2D section of a transversely isotropic material where $\Gamma$ is isotropic in the plane of the platelets and is symmetrical about the axis perpendicular to the platelets with maximum $\Gamma(0)=\Gamma_{\perp}$ and minimum $\Gamma(\pi/2)=\Gamma_{\parallel}$.
Polar plots of $\Gamma$ for different anisotropy values are shown in \Fig~\ref{fig:anisotropy-function}.

\begin{figure}[!htb]
\begin{center}
\includegraphics[width=\columnwidth]{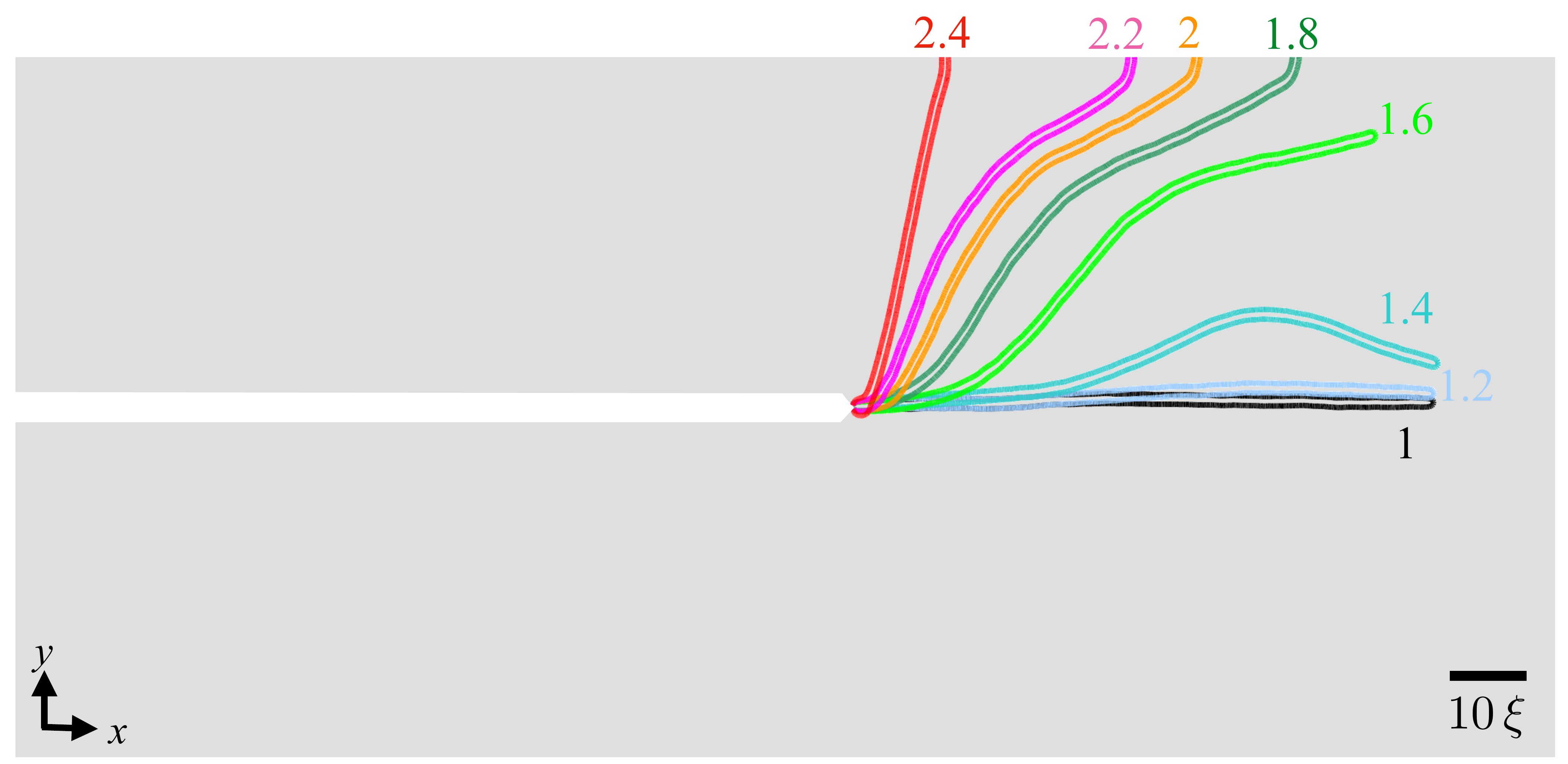}
\caption{
Phase-field fracture paths for $\xi=225\,\micron$ (\underline{mirrored to $y>0$}) showing a smooth transition from straight to kinked crack propagation in long samples with increasing fracture energy anisotropy.
}
\label{fig:crack-path-pfm}
\end{center}
\end{figure}

Fracture simulations were conducted for the same geometries studied experimentally (\Fig~\ref{fig:figure1} left column) varying $\A$ over the range (1 to 5) determined from experimental measurements of the fracture energy anisotropy around the kinking transition (\Fig~\ref{fig:angle-exp}).
The phase-field simulations results are in remarkably good quantitative agreement with experiments.
{The results of the phase-field simulations show that, in absence of fracture energy anisotropy the crack propagates straight in both samples.
This is not surprising since both sample geometries and load configurations are symmetric with respect to the horizontal axis and create no mode-II stresses.
}
Consistent with the crack paths shown in \Fig~\ref{fig:crack-path-exp}, simulations show that crack propagate straight in short samples (see \Fig~\ref{fig:rotated}b), but exhibit a smooth transition from straight to kinked propagation in long samples (\Fig~\ref{fig:crack-path-pfm}) with increasing anisotropy.
{In particular cracks in long sample for $\A>2.4$ (see \Fig~\ref{fig:crack-path-pfm}) kink sharply upon propagation.
The transition from straight propagation to kinked can be understood intuitively in terms of competition between the ability of the crack to release the stored elastic energy and the energetic cost of creating a new surface growing a crack.
Since the maximum normal stresses in the specimen are oriented in the y-direction, a crack propagating along the x-axis would release the highest amount of energy.
However, the energetic cost of propagating perpendicular to the platelets increases with increasing $\A$ ($f_v$).
Therefore, the kinked crack path provides a compromise to balance the cost of propagating in an ``easy direction'' with smaller amount of energy released.}

\begin{figure}[!htb]
\begin{center}
\includegraphics[width=\columnwidth]{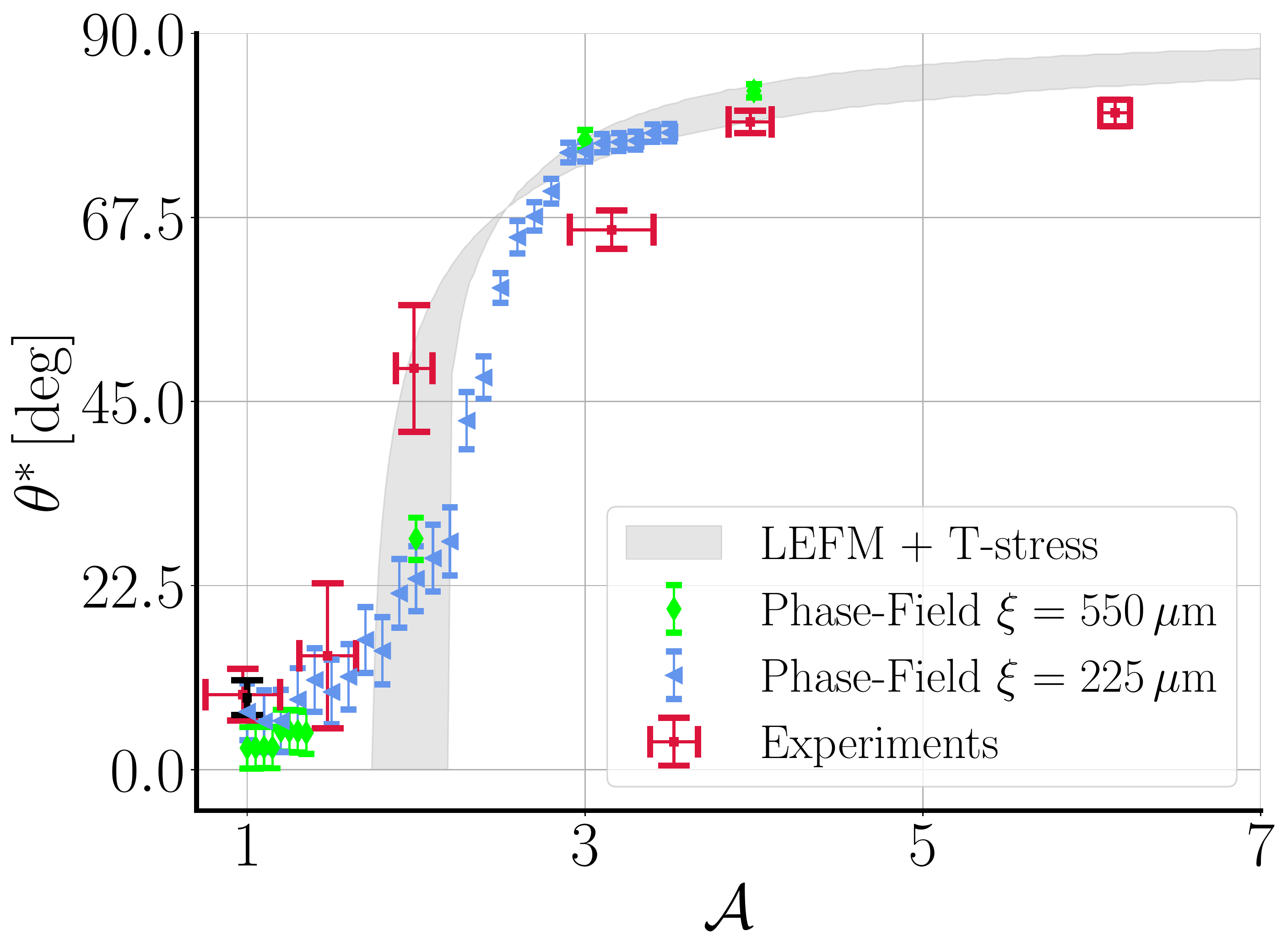}
\caption{
Comparison of simulated and experimentally observed initial kink angles in long samples. Experiments are for different volume percentages of platelets: $f_v=0$ (black) and $f_v=0.01\textup{--}0.07$ (red) where horizontal and vertical error bars signify the standard error of fracture energy anisotropy (\Fig~\ref{fig:exp-measurements}d) and initial kink angle (\Fig~\ref{fig:angle-exp}), respectively. Simulation results are shown for different anisotropies with the associated standard deviation due to discretization for $\xi=550\,\micron$ (green triangles)  and $\xi=225\,\micron$ (blue diamonds). Also shown for comparison are the LEFM predictions with T-stress corrections Eq.~\eqref{eq:kink-angle} (shaded gray area corresponding to $225 \le \xi \le 550\,\micron$ and $0.06\le T\sqrt{\xi}/K_I\le 0.09$) for the long sample geometry (see \Fig~\ref{fig:sT-long-short}). 
}
\label{fig:angles-anisotropy}
\end{center}
\end{figure}

This transition is further quantified in \Fig~\ref{fig:angles-anisotropy} where we plot the initial kink angle $\theta^*$ as a function $\A$ and superimpose experimental measurements of $\theta^*$ using \Fig~\ref{fig:exp-measurements}d to relate $f_v$ and $\A$. 
{This remarkable agreement between the experimental measurements and the phase-field simulations (with isotropic elasticity) establishes the role of fracture energy anisotropy in deflecting the crack.
However, the question remains, why the cracks in the short sample propagate straight?}

\subsection{Sample geometry effect}\label{ssec:sample-geometry-effect}

{The effect of the sample geometry on crack kinking in different isotropic specimens has been long observed~\cite{Cotterell:1965,Cotterell:1970,Smith:2001,Smith:2006,Ayatollahi:2016} where the crack can kink due to the effect of non-singular stresses around the crack tip.
In their classic paper, Cotterell and Rice~\cite{Cotterell:1980} revealed the critical role of the sign of the T-stress (see~\eqref{eq:crack-tip-sxx}) on crack path stability by analyzing the smooth trajectory of a curvilinear crack initially perturbed by a small kink angle. 
Their analysis shows that trajectories deviate exponentially away from the parent crack direction for $T>0$ (unstable propagation) or return parabolically to this direction for $T<0$ (stable propagation). 
One limitation of this calculation is that it is conducted in a traditional linear elastic fracture mechanics (LEFM) framework that neglects the role of the process zone scale and thus cannot predict the dependence of crack kinking behavior on sample geometry observed here in the experiments and phase-field simulations. This strong geometrical effect is reflected in the fact that crack kink in long samples but not short ones for identical  As we show below, however, this dependence can still be predicted in a LEFM framework using the analysis of Amestoy and Leblond that includes a T-stress correction to the energy release rate at the tip of a kinked crack \cite{Amestoy:1992}. This relative magnitude of this correction is proportional to the ratio $T\sqrt{\xi}/K_I$ of non-singular ($\sim T$) and singular ($\sim K_I/\sqrt{\xi}$) stresses on the process zone scale. 
A main finding here of the present work is that this ratio turns out to be sufficiently important to affect crack kinking behavior even when the process zone size $\xi$ is one to two orders of magnitude smaller than the sample size, as in the composites used here where sample sizes are on the cm scale (\Fig~\ref{fig:schematics}) and $\xi$ is estimated to be on the scale of a few hundred microns (Eq. \eqref{eq:process-zone}).

Similar to our results, Ayatollahi et al.~\cite{Ayatollahi:2016} showed that for double cantilever beams (DCB) and compact tension (CT) specimens made of (isotropic) PMMA, stability of cracks is affected by the specimen size and geometry through the sign and magnitude of T-stress.
In their study, they showed that the crack stability is well predicted by the dimensionless T-stress value $T\sqrt{\xi}/K_I$ for an appropriate choice of the process zone size $\xi$.
For the present experiments in anisotropic media, crack kinking can be predicted quantitatively by phase-field modeling, which phenomenologically regularizes stress field divergences on a scale $\xi$ determined from materials properties and hence inherently captures T-stress effects.


{Let us now turn to the analysis of crack kinking in the LEFM framework using analytical expressions for the energy release rate at the tip of a kinked crack that explicitly take into account T-stress effects~\cite{Amestoy:1992}.
For this, we consider a short extension of the parent crack of length $s$ and kink angle $\theta$ with respect to the parent crack axis. 
To predict $\theta$, we use the common assumption ~\cite{he1989kinking,ahn1998criteria,hakim2005crack,Hakim:2009,Takei:2013} that cracks propagate in a direction that maximizes the rate of decrease of the total energy, which is the difference $G(\theta)-\Gamma(\theta)$ (see Eq.~\eqref{eq:anisotropy}) between the energy release rate and the rate of increase of fracture energy. 
We use known analytical expressions for the stress-intensity-factors (SIFs) at the tip of a kinked crack that take in account the contribution of the T-stress~\cite{Amestoy:1992} to compute SIFs at the tip of the kinked crack. 
Given the stress intensity factor $K_I$ and T-stress $T$ of the parent crack (see~\eqref{eq:crack-tip-sxx}--\eqref{eq:crack-tip-syy}) and the fact that $K_{II}\equiv0$ by symmetry, the SIFs at the tip of the kinked crack can be written in the form
\begin{align}
	k_I(\theta)&=K_I f_{11}(\theta)+T\sqrt{s}\,g_1(\theta)\\
	k_{II}(\theta)&=K_I f_{21}(\theta)+T\sqrt{s}\,g_2(\theta)
	\label{eq:Amestoy-Leblond}
\end{align}
where the analytical expressions for the function $f_{ij}(\theta)$ and $g_i(\theta)$ are in Ref.~\cite{Amestoy:1992}. 
To calculate the energy release rate at the tip of the kinked crack $G(\theta)$, we use the standard expression~\cite{lawn1993fracture} 
\begin{equation}
	G(\theta)=\frac{1}{E}\left({k_I(\theta)^2}+{k_{II}(\theta)^2}\right)
\end{equation}
valid for plane stress. 
Given the relative magnitude of the non-singular to singular stresses $T\sqrt{s}/K_I$, we need to determine simultaneously the kink angle $\theta^*$ and the load $G(0)/\Gamma_{\perp}=K_I^2/(E\Gamma_{\perp})$ on the parent crack at which kinking occurs, where $\Gamma_{\perp}$ is the reference fracture energy at onset of straight propagation at $\theta=0$. Those two unknowns are determined by requiring that the angle that corresponds to a maximum of $G(\theta)-\Gamma(\theta)$ plotted versus $\theta$ in the range $0<\theta<\pi/2$ also satisfies the condition $G(\theta^*)=\Gamma(\theta^*)$ corresponding to the onset of propagation of the kinked crack. Those two conditions can be written in the succint form
\begin{equation}\label{eq:kink-angle}
	\theta^*=\underset{
						\begin{array}{c}
							0<K_I^2/E\leq \Gamma_{\perp}\\
							T\sqrt{s}/K_I~\mathrm{fixed}
						\end{array}
						}{\argmax}\left(G(\theta)-\Gamma(\theta)\right)
\end{equation}
When crack kinking does not occur ($\theta^*=0$), crack propagation occurs at a load corresponding to straight propagation of the parent crack $K_I^2/E= \Gamma_{\perp}$. In this case $G(\theta)-\Gamma(\theta)$ first crosses zero from negative to positive values at a maximum located at $\theta^*=0$.  In contrast, when kinking occurs $G(\theta)-\Gamma(\theta)$ first crosses zero at a maximum located at $\theta^*\ne 0$ and, concomitantly, crack propagation occurs at a lower load $K_I^2/E< \Gamma_{\perp}$ since straight propagation is energetically forbidden ($G(0)<\Gamma(0)$) in this case. 
We expect physically that the crack extension length used for the purpose of calculating the energy release rate at the tip of a nascent kinked crack should scale like the process zone size, or $s\sim \xi$. For the purpose of making a quantitative prediction, we choose $s=\xi$. Even though this choice is arbitrary up to a numerical prefactor of order unity, we have found that it predicts quantitatively well the effect of sample geometry on crack kinking behavior in both experiments and phase-field simulations. 
}

{ To quantify the T-stress effect in long and short samples, we numerically computed the T-stress from an analysis of stress fields in the vicinity of the crack tip (see~\ref{ssec:T-stress-approx}).
As previously shown in \Fig~\ref{fig:sT-long-short}, these computations yield $-0.20\le T\sqrt{\xi}/K_I\le -0.13$ in the short sample and $0.06\le T\sqrt{\xi}/K_I\le 0.09$ in the long sample for the process zone size $\xi$ in the estimated range $225\textup{--}550\,\micron$. 
For the long sample, this analysis (LEFM + T-stress) predicts that the positive T-stress is destabilizing and shifts the kinking transition to smaller fracture energy anisotropy than predicted by LEFM with vanishing T-stress (\Fig~\ref{fig:angles-anisotropy}), in good quantitative agreement with both experimental and phase-field simulation results.
For the short sample geometry, the analysis predicts that the negative T-stress has a stabilizing effect and shifts the kinking transition to larger anisotropy consistent with the absence of kinking in short samples over the same range of $\A<5$.}

{We note that $S(\theta)\equiv \Gamma(\theta)+d^2\Gamma(\theta)/d\theta^2$ is positive for all $\theta$ for the form assumed in the modeling. Consequently, the onset of kinking cannot be simply interpreted here using the criterion $S(\theta)<0$ that has been proposed to interpret forbidden crack propagation directions in tearing experiments ~\cite{Takei:2013}, where $G(\theta)$ is geometrically determined and has a different functional form than in the present 2D plane stress configuration. In addition, this criterion neglects the role of the process zone size shown here to strongly influence the critical anisotropy for kinking.}

Our experimental, numerical, and analytical results thus demonstrate that, unlike geometries considered by Ayatollahi et al.~\cite{Ayatollahi:2016}, in the long sample geometry the geometry effect alone is not enough to destabilize straight propagation.
However, the added effect of fracture energy anisotropy can result in the deflection of the crack.
On the other hand, while our short sample geometry stabilizes the crack path for small fracture energy anisotropies, we can hypothesize that cracks get deflected for large enough $\A$.
We tested this prediction by repeating a fracture experiment in a short sample for a higher volume fraction ($f_v=0.1$ corresponding to $\A\simeq9$) and phase-field simulations. 
Both experiment \Fig~\ref{fig:short-sample-kink}a and simulations \Fig~\ref{fig:short-sample-kink}b produced kinking as predicted, albeit the kinking occurred at a lower threshold in simulations $A\simeq5.75$ compared to the experiments.
We attribute the source of this discrepancy to the approximate form of the fracture energy anisotropy function $\Gamma(\theta)$ used.
Despite this discrepancy, we can conclude that the difference in the observed crack paths between long and short sample geometries stems  in phase-field simulations from the combination of the process zone size and the sign and magnitude of the T-stress in the two samples. 
The results suggest that the same combination is controlling crack path selection in the experiments even though the fracture processes on the process zone scale are only phenomenologically modeled in the phase-field approach. 

\begin{figure}[htb!]
\begin{center}
\includegraphics[width=\columnwidth]{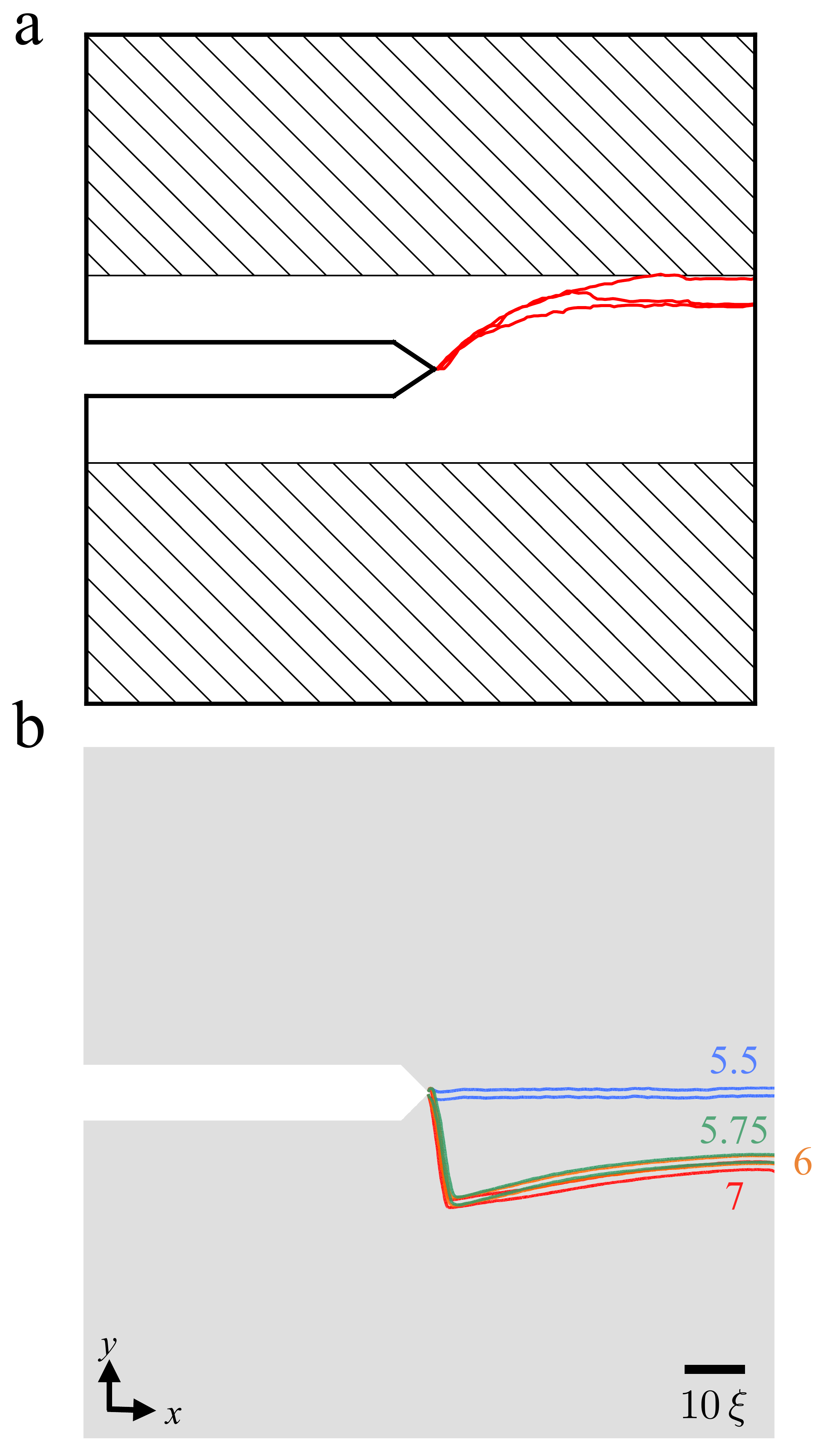}
\caption{Crack kinking in the short sample.
\textbf{a}, Experimental crack path in short sample geometry at 10\% volume fraction ($\A\simeq9$) showing crack kinking.
\textbf{b}, Results of phase-field simulation in short sample showing transition to kinking at $\A\simeq5.75$.} 
\label{fig:short-sample-kink}
\end{center}
\end{figure}

\subsection{Crack kinking for other platelet orientations}

\begin{figure}[htb!]
\begin{center}
\includegraphics[width=\columnwidth]{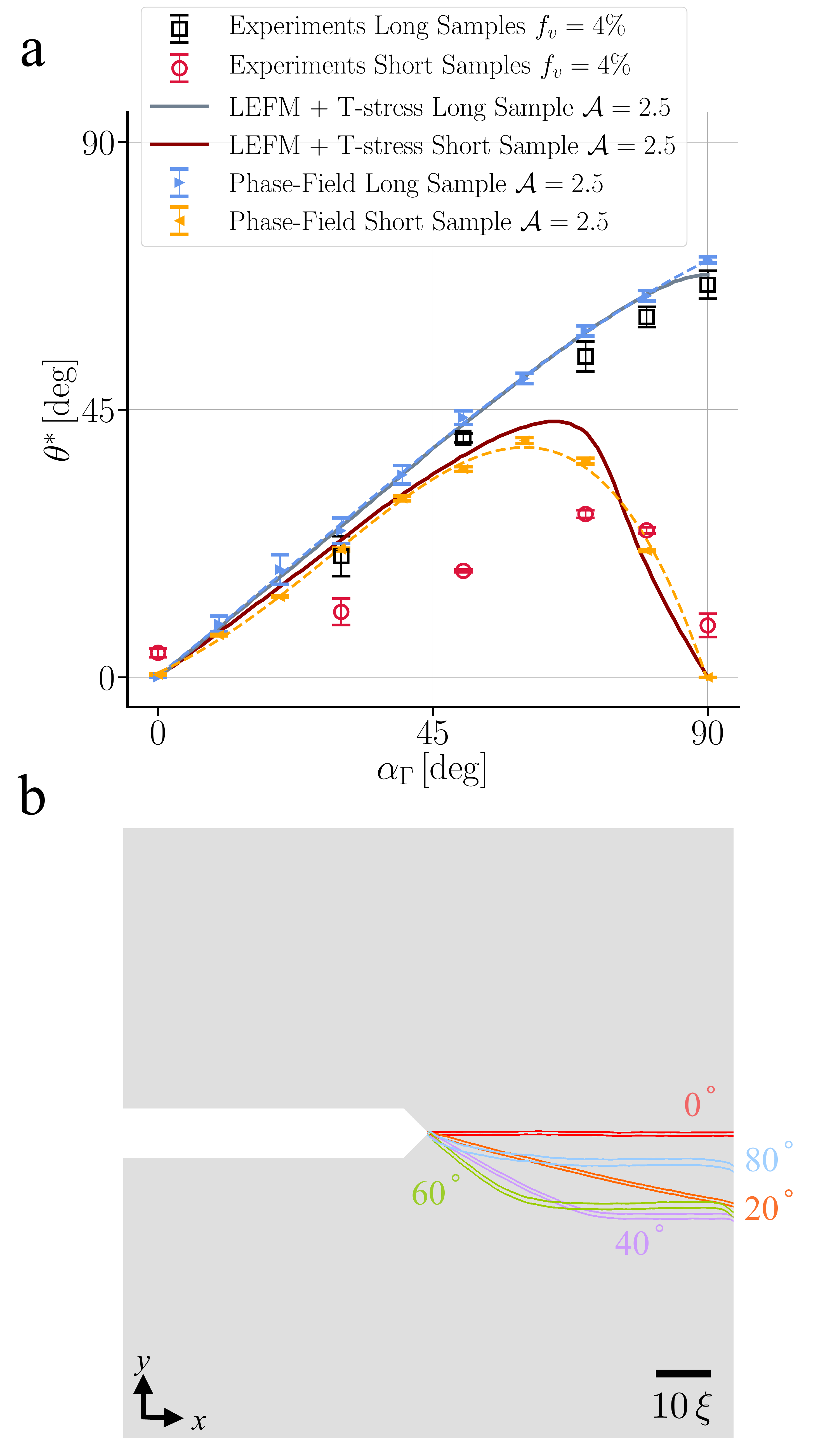}
\caption{
\textbf{a}, Comparison of experimental measurements, phase-field simulations, and LEFM theory + T-stress for the kink angle $\theta^*$ as a function of the angle $\alpha_{\Gamma}$ between the materials axis corresponding to the lowest fracture energy and the $x$-axis of the initial crack.
The comparison is shown for both short and long 4\% ($f_v=0.04$) samples to highlight the role of the T-stress on crack kinking.  
In this plot, $\alpha_{\Gamma}=0^\circ$ and $\alpha_{\Gamma}=90^\circ$ correspond to the $\parallel$ and $\perp$ orientations, respectively. 
The T-stress is seen to have a strong influence on kinking as seen by the large difference of kink angles in short and long samples for a wide range of angles larger than approximately $\alpha_{\Gamma}=45^\circ$.
Two values of fracture energy anisotropy $\A=2.5,3$ where used to assess the effect of the anisotropy on kinking.
\textbf{b}, Illustration of crack paths from phase-field simulation results for different $\alpha_{\Gamma}$ values in the short samples.}
\label{fig:rotated}
\end{center}
\end{figure}

To investigate the effect of sample geometry on crack paths for other orientations than the $\parallel$ and $\perp$, we carried out a series of experiments and phase-field simulations with varying platelet orientation (i.e. varying angle $\alpha_\Gamma$ of platelets with respect to the horizontal axis) between 0 and 90 degrees in both the short and long sample geometries. We chose a 4\% volume fraction of platelets ($f_v=0.04$) for which cracks propagate straight in short samples but kink in long samples for $\alpha_\Gamma=90^\circ$ due to the sample geometry effect.
{We also performed additional 2D plane stress phase-field simulations in both geometries by rotating the fracture energy anisotropy by an angle $\alpha_\Gamma$.}
The results are reported in \Fig~\ref{fig:rotated} where we plot the kink angle $\theta^*$ as a function of $\alpha_\Gamma$  (\Fig~\ref{fig:rotated}a) and show some examples of computed crack paths in the short sample (\Fig~\ref{fig:rotated}b). 
The results in \Fig~\ref{fig:rotated}a show that the kink angle differs significantly between the short and long samples for angles larger than about $45^\circ$, thereby demonstrating that the sample geometry strongly affects cracks paths for a wide range of other orientations than the perpendicular one.
For  $\alpha_\Gamma<45^\circ$, the geometry effect is small and the crack path follows approximately the low fracture energy direction ($\theta^*\simeq \alpha_\Gamma$). 
{To obtain the analytical prediction of kink angles (LEFM + T-stress in \Fig~\ref{fig:rotated}), we extend Eq.~\eqref{eq:kink-angle} to maximize the net energy release rate at fixed $T\sqrt{s}/K_I$ and $\alpha_{\Gamma}$ in
the situation where the fracture energy anisotropy function $\Gamma(\theta)$ is rotated by an angle $\alpha_\Gamma$.  Eq.~\eqref{eq:kink-angle} becomes
\begin{equation}
	\theta^*=\underset{
						\begin{array}{c}
							0<K_I^2/E\leq \Gamma_{\perp}\\
							T\sqrt{s}/K_I~\mathrm{fixed}\\
							\alpha_{\Gamma}~\mathrm{fixed}
						\end{array}
						}{\argmax}\left(G(\theta)-\Gamma\left(\theta-\alpha_{\Gamma}+\frac{\pi}{2}\right)\right)
\end{equation}
where the above maximization is carried out similarly to the last section using $s=\xi$. The load $G(0)/\Gamma_{\perp}=K_I/(E\Gamma_{\perp})$ is increased until the onset of propagation at $\theta^*$ such that $G(\theta^*)=\Gamma(\theta^*-\alpha_\Gamma+\pi/2)$.
LEFM theory with the inclusion of the T-stress is seen in \Fig~\ref{fig:rotated}a to predict well the kink angle dependence on platelet orientation for different sample geometries.}

{As previously shown, the crack path propagates straight in the short sample for $\parallel$ and $\perp$ orientations. However, at intermediate angles $0<\alpha_\Gamma<\pi/2$ the symmetry of the sample versus the x-axis is broken and therefore cracks cannot propagate straight for these platelet orientations.
We can further explain the transition cycle from straight propagation to deflection and back, in the short sample geometry, using the phase-field simulations results in \Fig~\ref{fig:rotated}. 
We observe that initially for $\alpha_\Gamma<60^\circ$ the kink angle $\theta^*$ increases with the platelets orientation $\alpha_\Gamma$.
However, at larger $\alpha_\Gamma>60^\circ$ (see $80^\circ$ in \Fig~\ref{fig:rotated}b for example) the initial kink occurs at a smaller angle followed by a subsequent straight propagation where the crack path is stabilized as a result of the mode-II loading due to its vertical shift.
Simply put, mode-II stresses create a configurational force acting perpendicular to the crack axis that balances out the effect of fracture energy anisotropy which tends to turn the crack axis into a direction that minimizes the surface energy $\Gamma(\theta)$~\cite{Hakim:2009}.
The discrepancy of the kink angle $\theta^*$ between phase-field simulations and the experimental observation in the short sample highlights the importance of the anisotropy function.
In absence of additional measurements of fracture energy anisotropy for the intermediate angles, our choice of the anisotropy function is only a first reasonable estimate that we can partially corroborate by non-trivial prediction of the onset of kinking with increasing platelet volume fraction in long samples ($\A_c\simeq 2$).
In future studies, it may be possible to measure the fracture energy for other orientations than $\parallel$ and $\perp$. 
Those measurements, however, are difficult because cracks in short samples kink for platelet orientations intermediate orientations. 
Consequently, the fracture energy cannot be extracted directly from force-displacement curves. This difficulty could be potentially circumvented by imposing mode-II (for example by offsetting the initial notch vertically) to force the crack to propagate straight in a medium of tilted platelets or by using numerical calculations to infer the fracture energy from load-displacement curves for those orientations.}
\section{Conclusions}\label{sec:conclusion}

{Our experimental and numerical results presented in this article highlight the interplay between the fracture energy anisotropy and sample geometry in crack path selection.
In this article, we combined experiments and simulations to show that while the crack path remains complex at the microscale, it is controlled at the macroscale by an emergent fracture energy anisotropy.
We further, demonstrated both numerically and experimentally that the onset of crack deflection not only depends on the microstructure (\eg volume fraction and orientation of platelets) but also is strongly influenced by the nonsingular T-stress which is a function of the geometry and loading configuration.
Our numerical simulations presented show how the phase-field fracture method can be used in conjunction with the experimental measurement to  predict crack path in orientationally ordered composites.}

Furthermore, our results suggest that in a natural composite such as bone with a much higher volume fraction of platelets ($f_v\simeq 0.4$), aligned mineralized collagen should suffice to produce crack kinking for propagation perpendicular to fibers independently of the sample geometry as observed experimentally~\cite{Koester:2008}, and that straight propagation in pathological bone~\cite{carriero2014tough} is due to a dramatic reduction of fracture energy anisotropy (i.e. $\A<\A_c$) caused by fiber misalignment. 
From a materials engineering standpoint, the strength of polymers reinforced by discontinuous ceramic filler is generally predicted with shear-lag theory in the literature~\cite{Bonderer:2008}, which assumes that the matrix and filler will ultimately fail through yielding and not through the brittle fracture of cracks propagating from defects. 
This treatment infers that composites are flaw tolerant during failure, a feature that is in potential disagreement with the low fracture energy exhibited by many ceramic filled polymer systems. 
In the current material system, for example, 5\% samples have order of a millimeter critical flaw sizes, suggesting that the crack geometries tested in this work reside in the brittle fracture regime and that shear-lag theory would over predict their performance.
This understanding should help interpret fracture experiments in a wide range of composites. 

\section{Acknowledgements}
This research was supported by NSF grant CMMI-1536354.
The majority of the numerical simulations presented in this work were performed using resources of the Extreme Science and Engineering Discovery Environment (XSEDE) under the resource allocation TG-MSS160013. 
The remainder of the simulations benefited from computing time allocation on Northeastern University's Discovery Cluster at the Massachusetts Green High Performance Computing Center (MGHPCC).

\section{Author contributions}

A.M. and C.P. are co-first authors listed in alphabetical order.
A.K., S.J.S., and R.M.E. conceived the research. A.M. and C.P. carried out the computational and experimental studies, respectively. A.M. and A.K. carried out the theoretical analyses.
The paper was written with input from all authors.

\bibliographystyle{apsrev}
\bibliography{mesgarnejad_PRE_2019}

\end{document}